\newcommand{\mc}{\multicolumn}
\documentstyle{mn}
\begin{document}

\title[The 7CQ sample I: selection and radio maps]
{Quasars from the 7C Survey - I: sample selection and 
radio maps}

\author[Riley et al.]{Julia M.\,Riley$^{1}$, Steve Rawlings$^{2}$, 
Richard G.\,McMahon$^{3}$,\\
\\
{\LARGE  Katherine M.\,Blundell$^{2}$,
Philip Miller$^{1,4}$, Mark Lacy$^{2}$ \& Elizabeth M.\ Waldram$^{1}$}\\
\\
$^1$Mullard Radio Astronomy Observatory, Astrophysics Group, Cavendish Laboratory, Madingley Road, Cambridge, CB3 0HE.\\
$^2$Astrophysics, Department of Physics, Keble Road, Oxford, OX1 3RH. \\
$^3$Institute of Astronomy, Madingley Road, Cambridge, CB3 0HA. \\
$^4$Present address: Venerable English College, Via di Monserrato, 45, 00186 Roma, Italy. \\
}
     
\maketitle

\begin{abstract}

We describe the selection of candidate radio-loud quasars obtained 
by cross-matching
radio source positions from the low-frequency (151 MHz) 7C survey
with optical positions from five pairs of $EO$ POSS-I plates scanned with
the Cambridge Automatic Plate-measuring
Machine (APM). The sky region studied is centred at RA 10$^{\rm h}$ 28$^{\rm m}$, 
Dec +41$^{\circ}$ and covers  $\approx 0.057 ~ \rm sr$.
We present VLA observations of the quasar candidates, and tabulate various
properties derived from the radio maps. We discuss the selection criteria of the
resulting `7CQ' sample of radio-loud quasars. The 70 confirmed quasars, 
and some fraction of the 36 unconfirmed candidates, constitute a filtered sample 
with the following selection criteria: 151-MHz flux density 
$S_{151}  > 100 ~ \rm mJy$; POSS-I $E-$plate magnitude $E \approx R < 20$; and
POSS-I colour $(O - E) < 1.8$; the effective area of the survey
drops significantly below $S_{151} \approx 200 ~ \rm mJy$. 
We argue that the colour criterion excludes few if any
quasars, but note, on the basis of recent work by Willott et al. (1998b),
that the $E$ magnitude limit probably excludes more than 50 per cent of the
radio-loud quasars.

\end{abstract}


\section{Introduction}
\label{sec:intro}

The advent of fast plate-measuring machines and extensive deep radio surveys 
with good positional accuracy has made possible the optical identification of large numbers 
of radio sources. Statistical studies can now be carried out using samples of radio sources
associated with particular classes of optical object such as quasars. Surveys at low radio frequency 
allow the selection of samples of quasars which avoid the strong orientation biases
inherent in samples selected at high radio frequencies: this is 
because the ratio of extended optically-thin radio emission to compact optically-thick radio 
emission increases rapidly with decreasing observing frequency, and it is the latter component which
is greatly enhanced by Doppler boosting when radio-emitting material moves 
at a relativistic speed along a direction close to the line-of-sight. Thus, although
low-frequency samples still contain some quasars whose total radio fluxes are 
strongly influenced by Doppler (i.e. orientation) effects, they are
far out-numbered by objects for which the optically-thin (steep-spectrum) component dominates ---
we will refer to the latter as steep-spectrum radio-loud quasars (SSQs).

In Fig.\ 1 we show how recent
work has extended the coverage of the low-frequency (151 MHz) radio 
luminosity\footnote{We take ${\em H}_{\circ}~=~50~{\rm
km~s^{-1}~Mpc^{-1}}$, ${\em q_{\circ}}~=~0.5$ and a cosmological constant of zero.}
$L_{151}$ {\it versus} redshift $z$ plane for SSQs. 
This provides a large range in $L_{151}$ at a given $z$ 
and {\it vice versa} --- these are important requirements if 
the influences of $L_{151}$ and $z$
on the properties of SSQs are to be separately determined.
However, no large samples selected at 151-MHz flux densities ($S_{151}$) 
below $\sim 0.5 ~ \rm Jy$ have until now been constructed. As Fig.\ 1 shows, such samples would 
improve the $L_{151}$--$z$ coverage still further. 
This provided the primary motivation for the project described in this paper.

\begin{figure*}
\begin{center}
\setlength{\unitlength}{1mm}
\begin{picture}(150,150)
\put(200,0){\includegraphics{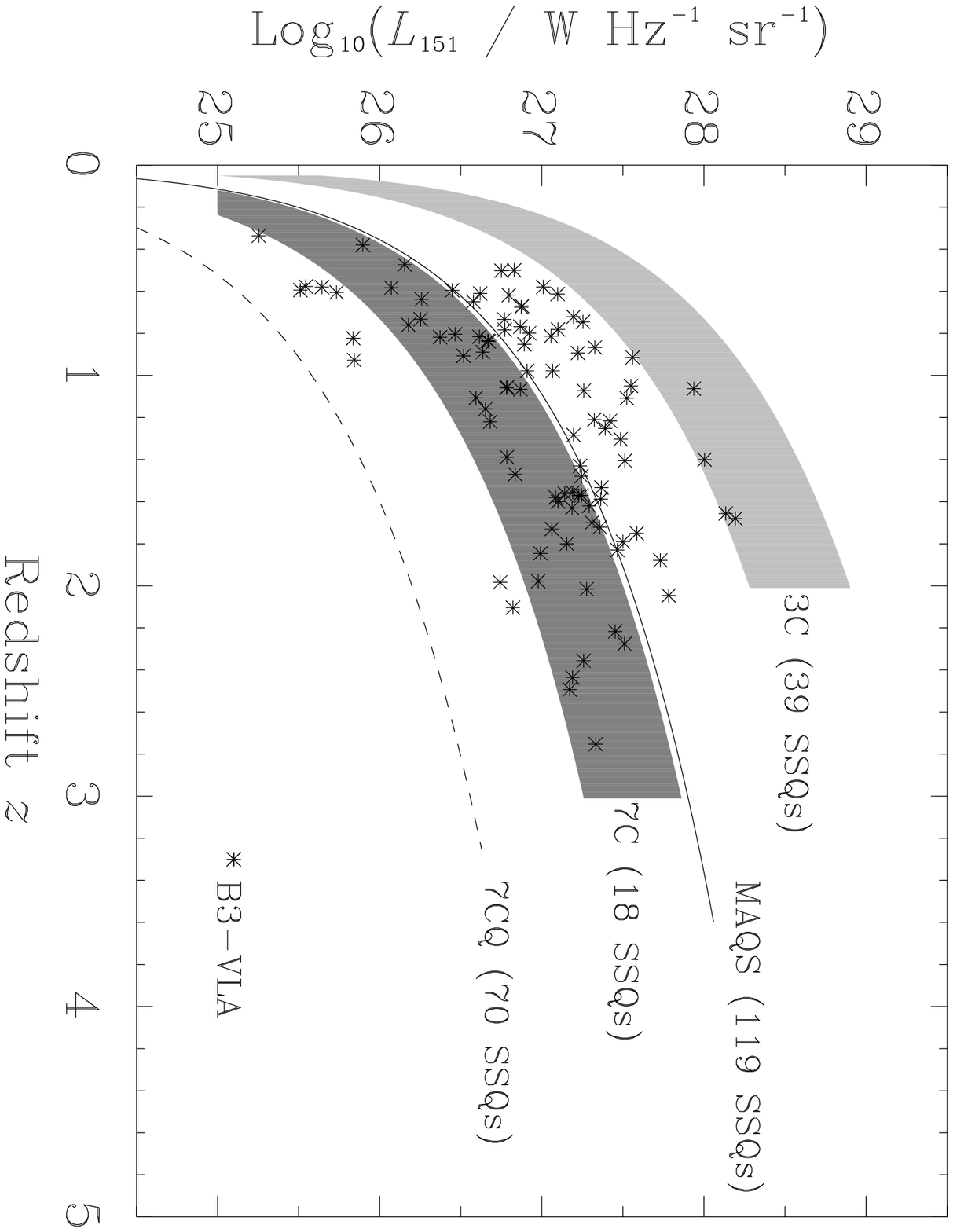}}
\end{picture}
\end{center}
{\caption[junk]{\label{fig:pz} The coverage of the 
$L_{151}$ {\it versus} $z$ plane for various samples of SSQs.
The location of SSQs within the 3C and 7C complete redshift surveys are shown by the
shaded regions; the lower bounds of these regions correspond to
$S_{178} = 10.9 ~ \rm Jy$ and $S_{151} = 0.5 ~ \rm Jy$ in each case taking
$\alpha = 0.8$. 
SSQs from the Molonglo/APM Quasar Survey
(MAQS), and also from the similarly-selected
Molonglo Quasar Survey (MQS), lie above the solid line
(which corresponds to a 408-MHz flux density limit of $0.95 ~ \rm Jy$
taking $\alpha = 0.8$). SSQs from the
B3-VLA survey are shown by asterisks.
Note that the B3-VLA, MAQS and MQS
surveys are all based on radio catalogues selected at 408 MHz.
Details of all these samples can be found in the following references: the
revised 3C sample -- Laing, Riley \& Longair (1983); 
the 7C Redshift Survey -- Willott et al. (1998b), and in prep.; 
the MAQS -- Serjeant et al. (1998), and in prep.; the MQS (Kapahi et al. 1998);
the B3-VLA survey -- Vigotti et al. (1997). 
SSQs from the 7CQ sample (described in this paper) lie 
above the dashed line which corresponds to 
$S_{151} = 0.1 ~ \rm Jy$ taking $\alpha = 0.8$. 
}}
\end{figure*}

The 7C surveys carried out with the Cambridge Low Frequency Synthesis Telescope (CLFST) 
at 151 MHz provide a perfect basis for this project because of their low limiting flux density,
$S_{151} \sim 0.1 ~ \rm Jy$ (e.g., McGilchrist et al. 1990; Lacy et al. 1995). 
Moreover, these surveys have a positional accuracy of a few arcsec and can be used to
identify candidate quasars by comparison with optical plates.
We chose to exploit these features by
cross-matching the 7C positions with the positions of optical objects from the
Palomar Observatory Sky Survey (POSS-I). We used the Automatic Plate-measuring 
Machine (APM) at Cambridge to scan the glass copies 
of the original Palomar Observatory Sky Survey (POSS-I) covering a continuous
area of the 7C survey;
in the search for quasar candidates we sought identifications (IDs) on both the
$E$ and $O$ POSS-I plates to minimise any selection biases due to object 
colour.

The organisation of this paper is as follows. In Sec.~\ref{sec:sample} we describe the selection of the
candidate radio-loud quasars. In Sec.~\ref{sec:VLA} we describe the VLA observations of 134 
of these candidates and in Sec.~\ref{sec:results} we present the 
results of these observations.
In Sec.~\ref{sec:discussion} we discuss the causes of incompleteness in the resulting sample of 
radio quasars which we will henceforth refer to as the 7CQ sample. In a 
companion paper (Rawlings et al., in prep.; hereafter Paper II) the optical spectra of the 
quasar candidates will be presented, and further properties of the sample discussed.

B1950.0 co-ordinates are used, and the convention for radio spectral index 
($\alpha$) is that $S_{\nu} \propto \nu^{-\alpha}$, where
$S_{\nu}$ is the flux density at frequency $\nu$.

\section{Sample definition}
\label{sec:sample}

We chose one area of the CLFST 7C survey centred at RA 
10$^{\rm h}$ 28$^{\rm m}$ and Dec +41$^{\circ}$ (McGilchrist et al. 1990)
for our new study of SSQs; a typical value for the 
limiting ($5 \sigma$) flux density over this area is $80 ~ \rm mJy$ (for point sources) but in
some parts it is as high as 
$\approx 200 ~ \rm mJy$, and the $1 \sigma$ positional 
errors range from $\sim$2 arcsec for sources with flux densities $> 1 ~ \rm Jy$ to $\sim$10 arcsec for the 
faintest sources.

Five pairs of POSS-I plates were scanned ($EO$690, $EO$709, $EO$711, $EO$1032 and 
$EO$1348).  A catalogue was constructed of all the optical objects with stellar images 
on both the $E$ and $O$ plates which were within 15 arcsec of a radio source; there were 
175 such objects, of which 7 were on two adjacent POSS-I plates.  A statistical 
analysis described in Sec.~\ref{sec:discussion}
indicated that a significant fraction of these objects are random coincidences, 
particularly those for which the radio-optical separation is greater than 10 arcsec.  

We decided therefore to concentrate follow-up work on those objects with 
radio-optical separations ($r$) of less than 10 arcsec.  Further higher resolution radio 
observations were required to eliminate the remaining chance associations.  We 
therefore made VLA observations of those sources in the sample with 
$S_{151} > 0.1 ~ \rm Jy$, and $r < 10 ~ \rm arcsec$; due to pressure on observing time 
a small fraction ($\sim 10$ per cent) of the sources were omitted. Some sources were omitted either because they were very 
extended at 151 MHz ($> 40 ~ \rm arcsec$), or because they had already been observed with the VLA.
Other sources were omitted randomly if there
were too many targets in a particular RA range during the VLA observations.
Another source of essentially random exclusion arose because of 
minor changes in sample membership to be discussed in the following paragraph.
We also chose to eliminate some sources
from our VLA mapping programme because of their red (optical) colours; the
reasons for this are discussed in Sec.~\ref{sec:discussion} (these
sources were mainly associated with the $EO$709 POSS-I plate pair).  
Some of the sources with $r$ in the range 10--15 arcsec were also observed 
with the VLA if the optical object seemed likely, on the grounds of colour, to be a quasar.

Subsequently the optical plates were re-scanned and the APM data re-calibrated.
This allowed for a number of minor improvements to the APM analysis procedures 
(which were then identical to those described by Lacy et al. 1995).
A new catalogue was constructed of all the optical objects which were stellar on either the $E$ 
or the $O$ plates, or on both, with $r < 10 \rm ~ arcsec$; it contains 206 entries of which
six were repeated measurements of an optical object on
two pairs of plates, and four were radio sources each matched with two possible optically-unresolved
IDs (we excluded the two radio sources found serendipitously in one field, category `T' in Table 1, 
since they fail the $r < 10 ~ \rm arcsec$ criterion). There are
thus 196 possible 7C/APM cross matches in this revised main sample. There are fairly minor differences between this 
and the initial catalogue which relate to re-classification of some of the optical objects.
Table 1 contains the main sample and was constructed from 
the second catalogue since this is more reliable (sources not in the first catalogue 
are flagged). Table 2 lists a subsidiary sample consisting of
26 additional cross-matches which were from the first version of the catalogue either with $10 < r < 15 ~
\rm arcsec$ (and with the blue colours expected for quasars)  or with $r < 10 ~ \rm arcsec$ 
but not in the second catalogue (due to optical re-classification).

The total sky coverage of the 7CQ sample is the area of the five POSS-I plate pairs, or about
$0.057 ~ \rm sr$. The limiting optical magnitudes of the 
POSS-I plates are $O \sim B \approx 21.5$ and
$E \approx R \approx 20$, and the typical uncertainty in an $O - E$ colour is about
0.5 magnitudes. The reliability of the APM classification algorithm, i.e.
establishing whether objects are unresolved or extended, varies 
with object magnitude. Only for objects about two magnitudes brighter than the plate limit 
is it possible to separate point-like optical objects from extended objects with very high reliability.
Close to the plate limit circular and/or small galaxies are indistinguishable
from point-like quasars. One of the objects, while selected in the initial
version of the catalogue, was not selected in the final version (and hence is listed only in 
Table 2) because of optical re-classification: it is now classified by the APM analysis as
extended, but is a spectroscopically-confirmed high-redshift quasar
(Paper II). There is thus a small risk of falsely excluding point-like 
optical objects at the fainter
optical magnitudes. These issues will be discussed further in Section 5.2.

\begin{figure*}
\label{fig:liz}
\begin{center}
\setlength{\unitlength}{1mm}
\begin{picture}(200,200)
\put(0,-30){\includegraphics{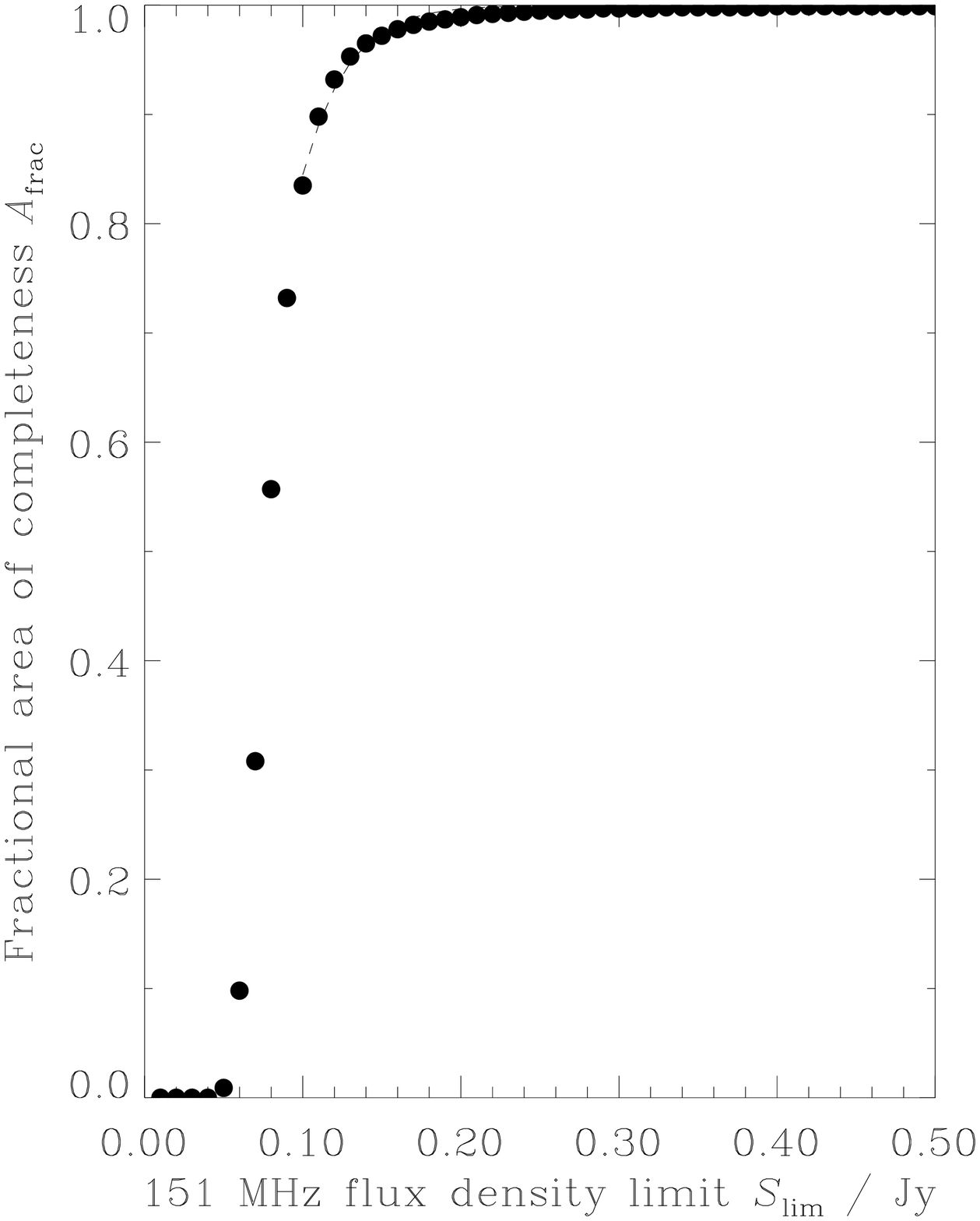}}
\end{picture}
\end{center}
{\caption[junk]{\label{fig:pz} The fraction $A_{\rm frac}$ of the 7CQ area which is
complete to a given 151 MHz limiting flux density $S_{\rm lim}$. The dashed line
shows the fit to the data at $S_{\rm lim} > 0.1 ~ \rm Jy$ discussed in Sec.~\ref{sec:sample}.
}}
\end{figure*}

The 151 MHz flux density completeness limit varies across each 7C field because of 
the primary beam polar diagram, geometric and other distortions of the beamshape 
(McGilchrist et al. 1990), and because of fluctuations in the local noise level.
We have calculated the fraction $A_{\rm frac}$ of the total 7CQ survey area of 
$0.057 ~ \rm sr$ which is complete to a given limiting flux density $S_{\rm lim}$
taking all these effects into account, and this is plotted in 
Fig.\ ~\ref{fig:liz}. At $S_{\rm lim} > 0.1 ~ \rm Jy$ a suitable functional
approximation to this curve is given by $A_{\rm frac} = (1 + e^{-a (\it S_{\rm lim} - b)})^{-1}$
with $a = 39.72$ and $b = 0.0573$. This approximate functional
form will be used in Paper II.

\clearpage

\scriptsize
\begin{table*}
\begin{center}
\begin{tabular}{|l|ll|r|r|r|r|r|r|r|r|r|r|r|l}
\hline\hline
(1) & (2) & (3) & (4) & (5) & (6) & (7) & (8) & (9) & (10) & (11) & (12) & (13) & (14) & (15) \\
\hline
& \mc{1}{c}{RA(B1950)} & \mc{1}{c|}{Dec(B1950)} & \mc{1}{c|}{$r$}
& \mc{1}{c|}{$O$} & \mc{1}{c|}{$O-E$} & \mc{1}{c|}{$\Delta \rm RA$} & \mc{1}{c|}{$\Delta \rm Dec$} 
& \mc{1}{c|}{$S_{151}$} & \mc{1}{c|}{$S_{1490}$} & \mc{1}{c|}{$\alpha_{151}^{1490}$} &
\mc{1}{c|}{$S_{\rm core}$} & \mc{1}{c|}{$\theta$} & \mc{1}{c|}{Opt} & 
\\
\hline
 & 09 55 32.61 & 43 30 40.9 & 2.1 & 20.64 &  1.39 & -0.07 &  1.5 &   690 &  174   & 0.60 &  174.0 &  $\leq0.3$  & QC  &  N \\
 & 09 55 49.88 & 42 51 25.9 & 4.2 & 18.13 &   .53 & -0.06 &  0.5 &   520 &   87   & 0.78 &   20.4 &       11.0  & QC  &  Y \\
 & 09 56 34.8  & 40 29 52.0 & 6.4 & 17.83 &  1.44 &  0.42 &  6.6 &  1080 & (118)  &(0.99)&        &        8.0  &  S  &  V \\
 & 09 58 07.4  & 42 44 07   & 9.4 & 21.63 &  2.49 &  0.09 & -9.8 &   180 &  (m)   &      &        &      7C(43) &  ?  &  N$^{r}$\\
 & 09 58 31.39 & 41 21 19.0 & 6.8 & 19.97 &  1.31 & -0.04 &  0.5 &   420 &   54   & 0.90 &    1.5 &       12.0  & QC  &  Y \\
 & 09 58 37.59 & 42 23 38.9 & 6.8 & 19.51 &  2.60 &  0.16 & -2.2 &   230 &   45   & 0.71 &   45.0 &        0.9  &  S  &  Y \\
 & 09 59 14.5  & 40 48 50   & 3.2 & 21.04 &  2.37 &  0.62 & -5.5 &   280 &   24   & 1.07 &        &       34.0  &  M  &  Y \\
 & 09 59 27.06 & 40 07 39.6 & 7.7 & 18.89 &  2.27 & -0.48 &-11.9 &   280 &   33   & 0.93 &   33.0 &        1.1  &  M  &  Y \\
N& 10 00 46.07 & 41 33 57.6 & 7.5 & 21.70 &  2.25 &  0.90 &-13.5 &   200 & (29.8) &(0.85)&        &       39.0  &  M  &  F \\   
 & 10 01 20.79 & 42 33 52.8 & 5.5 & 16.97 &  1.49 &  0.25 &  7.2 &   890 &   76   & 1.07 &        &        2.4  &  M  &  Y \\
 & 10 01 36.45 & 45 07 27.5 & 9.8 & 15.81 &   .98 & -0.16 &-11.8 &   130 &   14.4 & 0.96 &   14.4 &  $\leq0.3$  &  M  &  N \\
 & 10 01 39.73 & 40 51 21.6 & 6.2 & 20.95 &  1.70 & -0.03 &  0.0 &   130 &   52   & 0.40 &   39.0 &       17.0  &  Q  &  Y \\
 & 10 01 53.98 & 50 34 05.3 & 2.9 & 20.14 &  1.44 & -0.03 &  0.7 &   150 & (73.8) &(0.32)&        &     (19.0)  &OC0(c)& C \\
 & 10 02 05.90 & 48 20 08.7 & 2.8 & 17.57 &   .50 & -0.04 & -0.4 &   640 &  186   & 0.54 &  186.0 &        0.4  &  Q  &  N \\
N& 10 03 28.8  & 43 13 20.3 & 3.4 & 20.86 &  2.76 &  0.20 &  2.8 &   360 & (76.2) &(0.70)&        &     (70.0)  &EC1  &  C \\
\hline\hline
\end{tabular}
{\caption[Table 1]{\label{tab:tab1} {\tiny
A summary of the radio and optical properties of the 
candidate 7CQ quasars in the 
main sample. 
{\bf Column 1} contains flags with meanings as follows:
N indicates that the source was not in the first version of the
optical catalogue; R indicates a repeat entry because the optical object occurs in 2 POSS-I fields; 
S indicates that there is more than one optically-unresolved object in the field of the source; 
* indicates that this proposed identification failed the original selection criterion on $r$ (see column 4); and 
T indicates that there is more than one radio source in the field. 
The RAs and Decs of the radio sources are given in {\bf Columns 2 \& 3}: the positions were measured from the 
radio core if present (i.e., if there is an entry in column 12); 
where there is no core the position is midway between the 
radio hotspots, or the centroid of the radio emission; if the source has not been
observed at high resolution with the VLA, the position is taken from the NVSS 
catalogue (Condon et al. 1998) or (in a few cases marked 7C in column 13)
the 7C catalogue (McGilchrist et al. 1990).
{\bf Column 4:} values of $r$, the separation (in arcsec) of the 7C radio source
position and the APM optical position. {\bf Columns 5 and 6:} 
the magnitude of the quasar candidate on the POSS-I O-plate, and its colour 
(the difference between its magnitudes on the O- and E-plates).
{\bf Columns 7 and 8:} the differences in RA (in seconds of time) and Dec (in arcsec) 
between the radio position given in columns 2 and 3 and that of the quasar candidate 
(in the sense radio minus optical). {\bf Column 9:} 
the flux density in mJy at 151 MHz $S_{151}$ from the 7C catalogue. 
{\bf Column 10:} the flux density in mJy 
either at 1.49 GHz from our own VLA observations
or (in parentheses) at 1.40 GHz from the NVSS survey. In a few cases, marked (m), there is
no corresponding source in the NVSS survey and/or no source was detected in higher resolution
VLA observations to a typical surface brightness upper limit 
of $\sim 1$ mJy per beam on low-resolution (4 arcsec FWHM) maps made either from our
own observations or from the FIRST Survey data of Becker, White \& Helfand (1995).
{\bf Column 11:} the spectral index, $\alpha$, between 151 MHz and 1.49 (or 1.4) GHz; (c)
indicates that no value has been calculated because the value of
$S_{151}$ is severely over-estimated because of source confusion. 
{\bf Column 12:} the flux density of the core in mJy at 1.49 GHz.  
For well-resolved sources this is the height of the peak (in mJy per beam) at the core position.  
For less well-resolved sources in which it is impossible to distinguish 
the core from the surroundings it is the 
total flux density given in column 10. {\bf Column 13:} 
the angular size of the source in arcsec. A question mark flags an uncertainty which is 
is discussed in the notes on that soure. If the entries in columns 10 and 
12 are the same the size was measured by 
fitting a Gaussian with the {\sc AIPS} task {\sc IMFIT}.  
Otherwise the size was measured from the map. A value in parentheses is the largest angular size
given in the NVSS catalogue (or the 7C catalogue, if preceded by 7C, with a 'c'
denoting that the 7C catalogue entry is a confusion of two or more sources).
{\bf Column 14:} optical type, classified as follows: Q - spectroscopically-confirmed quasar (see paper II);
QC - good ID which is optically unresolved but which has not not yet been 
spectroscopically confirmed as a quasar or galaxy; S - proposed ID is a star;
M - proposed ID rejected on positional grounds; 
G - confirmed galaxy; BL - confirmed BL Lac object; ? - stellar on both the $O$ and $E$ plates but
ID not yet certain due to lack of follow-up with the VLA and optical spectroscopy;
EC1, EC2 - the object is non-stellar (1) or confused (2) on the $E-$plate; 
OC0, OC1, OC2 - the object is noise-like (0), non-stellar (1) or confused (2) on the 
$O-$plate. Note that some of the spectroscopically-confirmed quasars are 
not classified by the APM analysis as stellar on both POSS-I plates.
A (c) in this column indicates that the optical object is very likely to be an ID --
the criterion being that the NVSS and APM positions differ by $< 2 ~ \rm arcsec$ 
for unresolved radio sources, and $< 6 ~ \rm arcsec$ for resolved sources. 
{\bf Column 15} contains flags associated with the radio data:
a Y indicates that a map is presented in Fig.\ 3 with a lower case y indicating that additional
notes are to be found in the text; N indicates that a map was made but 
is not shown in Fig.\ 3 (typically because the source is unresolved) with lower case n,
again, advertising additional notes in the text; C indicates that the only map available is from the
NVSS low-resolution VLA survey of Condon et al. (1998); $\ddag$ means excluded from our high-resolution
VLA follow-up because $S_{151} \leq 100 ~
\rm mJy$, \makebox{$^{r}$ means} excluded because of the red optical colour of the proposed ID
(see Section 5), $^{l}$ means excluded because of large 7C angular size, and $^{j}$
means excluded randomly.
Other letters refer to additional sources of high-resolution radio maps:
D, Law-Green et al. (1995); 
F, FIRST survey archive (Becker et al. 1995); M, Machalski \& Condon (1986); L, Riley \& Warner (1994);
P, Patnaik et al. (1992); V, Vigotti et al. 1989. One entry in the 7C catalogue
(at position 10 24 04.2, +48 43 30) is an alias of a bright radio source, and its candidate identification
a star. }
}
}
\end{center}
\end{table*}

\normalsize

\addtocounter{table}{-1}

\clearpage

\scriptsize
\begin{table*}
\begin{center}
\begin{tabular}{|l|ll|r|r|r|r|r|r|r|r|r|r|r|l}
\hline\hline
(1) & (2) & (3) & (4) & (5) & (6) & (7) & (8) & (9) & (10) & (11) & (12) & (13) & (14) & (15) \\
\hline
& \mc{1}{c}{RA(1950)} & \mc{1}{c|}{Dec(1950)} & \mc{1}{c|}{$r$}
& \mc{1}{c|}{$O$} & \mc{1}{c|}{$O-E$} & \mc{1}{c|}{$\Delta \rm RA$} & \mc{1}{c|}{$\Delta \rm Dec$} 
& \mc{1}{c|}{$S_{151}$} & \mc{1}{c|}{$S_{1490}$} & \mc{1}{c|}{$\alpha_{151}^{1490}$} &
\mc{1}{c|}{$S_{\rm core}$} & \mc{1}{c|}{$\theta$} & \mc{1}{c|}{Opt} & 
\\
\hline
 & 10 03 55.4  & 38 27 56   & 8.1 & 19.57 &   .78 & -0.66 & -2.2 &   160 &  (m)   &      &        &     7C(71)  &  S  &  N \\
 & 10 05 31.10 & 45 33 23.3 & 6.4 & 19.66 &   .87 & -0.57 & -1.7 &   120 &   23   & 0.72 &        &        3.3  &  M  &  Y \\
N& 10 05 53.5  & 41 00 58   & 7.9 & 21.58 &  1.87 &  0.43 &  5.6 &   160 &  (m)   &      &        &             &EC1  &  F \\
 & 10 06 33.2  & 47 10 08   & 5.3 & 20.69 &   .80 &  0.14 & -2.8 &   450 &   77   & 0.77 &        &        7.0  &Q/EC1&  Y \\
 & 10 06 47.28 & 45 12 41.5 & 4.5 & 20.36 &  1.43 & -0.46 & 10.4 &    80 & (14.0) &(0.78)&        &      (23.9) &  ?  &  C$\ddag$\\
 & 10 07 44.81 & 36 36 36.6 & 4.5 & 21.38 &  1.77 & -0.10 &  3.7 &   160 &   23   & 0.85 &   23.0 &  $\leq0.3$  & QC  &  N \\
N& 10 07 52.96 & 34 24 41.7 & 1.3 & 15.29 &   .99 &  0.44 & 11.0 &    90 & (15.3) &(0.80)&        &       23.0  &  M  &  F$\ddag$\\
 & 10 07 57.46 & 33 45 05.2 & 4.5 & 19.98 &   .04 & -0.13 & -0.2 &   230 &  376   &-0.21 &  376.0 &  $\leq0.3$  &  Q  &  N \\
 & 10 07 59.11 & 38 32 23.0 & 4.1 & 19.91 &   .60 &  0.02 &  0.3 &   290 &   49   & 0.78 &   15.0 &       42.0  &  Q  &  Y \\
 & 10 09 17.53 & 33 24 16.2 & 4.2 & 17.99 &   .40 & -0.15 & -1.2 &   570 &(195)   &(0.48)&        &      ($<17$)&  Q  &  C \\
 & 10 09 21.46 & 40 44 24.2 & 3.3 & 21.75 &  2.42 & -0.01 & -0.8 &   130 &   13   & 1.0  &        &        3.5  & QC  &  y \\
 & 10 09 23.3  & 36 54 18   & 4.9 & 16.53 &  1.20 & -0.17 & -5.1 &   370 &   47   & 0.90 &        &       60.0  &  M  &  Y \\
 & 10 09 24.0  & 42 03 18   & 4.4 & 18.87 &  2.39 & -0.12 & -4.0 &   390 &  (m)   &      &        &     7C(65)  &  S  &  N \\
N& 10 09 42.36 & 42 44 48.2 & 1.1 & 18.13 &  1.56 & -0.08 &  0.8 &   180 & (79.5) &(0.36)&        &      ($<19$)&EC1(c)& C \\
 & 10 09 48.44 & 36 08 39.2 & 6.3 & 18.62 &   .89 & -0.04 &  0.1 &   180 &   25   & 0.85 &   25.0 &        0.6? &  Q  &  y \\
 & 10 09 54.24 & 38 34 33.4 & 6.0 & 21.24 &  3.43 & -0.14 & -1.4 &   360 &   54   & 0.83 &    2.4 &       32.0  &  S  &  Y \\
N& 10 09 58.96 & 46 10 31.7 & 5.4 & 21.74 &  3.53 &  0.10 &  2.7 &   160 & (46.1) &(0.56)&        &     ($<16$) &EC2  &  C \\
 & 10 10 20.74 & 49 33 33.5 & 3.6 & 19.35 &   .66 & -0.09 &  0.6 &   210 &  285   &-0.13 &  285.0 &  $\leq0.3$  &  Q  &  N \\
 & 10 10 32.9  & 47 48 49   & 8.6 & 19.50 &  1.59 &  0.81 &  0.0 &   110 & (m)    &      &        &         7C  &  ?  &  -$^{j}$ \\
 & 10 10 46.37 & 42 04 02.6 & 1.5 & 19.60 &   .94 &  0.02 &  1.1 &   500 &   72   & 0.85 &    2.1 &       21.0  & QC  &  Y \\
N& 10 11 13.89 & 40 08 07.5 & 7.7 & 19.58 &   .48 &  0.06 & -0.3 &   100 & (45.5) &(0.35)&   43.9 &       14.0  &EC1(c)&  F$\ddag$\\
 & 10 11 17.65 & 41 27 12.0 & 7.1 & 20.56 &  1.09 &  0.05 &  1.3 &   130 &   46   & 0.47 &   37.0 &        3.0  & QC  &  Y \\
 & 10 11 27.34 & 40 58 21.2 & 8.2 & 17.41 &  1.12 & -1.74 &  2.5 &   310 & (30.8) &(1.04)&        &       33.0  &  M  &  F \\
 & 10 11 44.58 & 44 36 28.4 & 1.2 & 18.96 &   .39 & -0.04 &  0.0 &   450 &   69   & 0.82 &    4.0 &       34.0  &  Q  &  Y \\
 & 10 11 55.26 & 49 40 56.4 & 1.9 & 16.47 &   .96 & -0.09 &  0.4 &   690 &  370   & 0.27 &  344.0 &        2.0  & BL  &  Y \\
 & 10 12 32.25 & 39 22 22.8 & 7.3 & 20.43 &  1.46 &  0.14 & -0.6 &   110 &   15   & 0.88 &   10.0 &        1.2  &Q/EC1  &  N \\
 & 10 12 47.1  & 32 42 02   & 7.9 & 20.71 &  1.77 &  0.58 & -4.1 &   910 &(110)   &(0.95)&        &       43.0  &  M  &  Y \\
 & 10 12 50.2  & 40 20 18   & 6.7 & 21.06 &  3.25 & -0.48 &-12.0 &   300 &   31   & 0.99 &        &       65.0  &  M  &  Y \\
N& 10 12 52.69 & 39 35 16.3 & 8.0 & 21.44 &  2.67 &  0.55 &-10.9 &   380 & (56.3) &(0.86)&        &       29.0  &  M  &  F \\
 & 10 13 21.83 & 36 24 35.7 & 4.7 & 18.42 &  2.74 &  0.34 &  2.4 &   640 &   61   & 1.03 &   47.0 &        2.0  &  M  &  Y \\
N& 10 13 24.60 & 46 58 41.1 & 8.4 & 19.75 &  2.56 & -0.54 & 4.5  &   530 & (137)  &(0.61)&        &     (27.0)  &  ?  &  C$^{j}$ \\
 & 10 14 18.96 & 46 34 11.5 & 7.0 & 20.41 &   .70 &  0.62 & -5.8 &   150 &   33   & 0.66 &   33.0 &  $\leq0.3$  &  M  &  N \\
 & 10 14 26.91 & 38 47 54.5 & 5.6 & 16.15 &   .94 &  0.43 & 14.4 &    80 &(20.8)  &(0.60)&        &       $<5$  &  S  &  F$\ddag$\\
 & 10 14 46.25 & 46 25 15.7 & 4.8 & 19.43 &   .51 & -0.07 & -0.6 &   230 &   42   & 0.74 &    3.8 &       14.0  &  Q  &  Y \\
N& 10 14 55.67 & 45 48 02.4 & 9.9 & 19.63 &  2.00 &  1.30 &  6.4 &   120 &(18.2)  &(0.85)&        &     (19.7)  &  ?  &  C$^{j}$ \\
 & 10 14 56.71 & 44 14 50.7 & 5.6 & 18.97 &   .72 &  0.02 &  1.2 &   110 &   24   & 0.67 &        &        3.0  &  Q  &  Y \\
 & 10 15 04.90 & 46 20 24.8 & 2.9 & 20.48 &  1.07 &  0.19 & -2.9 &   550 &   61   & 0.96 &        &        6.0  &Q/OC1&  Y \\
 & 10 15 16.23 & 35 57 41.3 & 3.6 & 18.64 &   .93 & -0.07 &  0.2 &   490 &  580   &-0.07 &  560.0 &        4.8  &  Q  &  Y \\
 & 10 15 36.56 & 43 37 34.6 & 1.4 & 20.69 &   .78 & -0.08 &  0.8 &   340 &   45   & 0.88 &   25.0 &        4.0  &  Q  &  Y \\
N& 10 16 35.86 & 33 53 14.3 & 9.8 & 21.42 &  1.68 &  0.16 & -8.6 &   100 & (7.7)  &(1.15)&        &    ($<34$)  &  ?  &  C$\ddag$\\
 & 10 16 41.74 & 42 59 01.4 & 3.1 & 19.76 &  1.18 &  0.24 & 14.7 &   160 & (25.8) &(0.82)&        &     (30.5)  &  S  &  C \\
 & 10 16 42.40 & 34 28 02.2 & 7.2 & 17.68 &  1.16 &  0.13 & 23.1 &    70 &  (7.7) &(0.99)&        &    ($<34$)  &  S  &  C$\ddag$\\
 & 10 16 49.79 & 38 48 35.7 & 5.6 & 18.56 &  2.38 & -0.12 & -2.0 &  1020 & (54.6) &(1.31)&        &       76.0  &  S  &  F \\
N& 10 17 19.48 & 46 11 41.7 & 0.5 & 21.34 &  3.36 & -0.06 & -1.2 &   120 & (36.8) &(0.53)&        &     (18.2)  &EC1(c)& C \\
N& 10 17 48.31 & 48 46 24.3 & 5.7 & 20.17 &  1.54 & -0.33 & 48.1 &  9300 & (10.3) &  (c) &        &    ($<23$)  &  M  &  C \\
 & 10 18 03.44 & 45 38 39.0 & 2.6 & 18.16 &   .50 & -0.02 & -0.6 &   630 &(131.1) &(0.70)&   47.0 &       45.0  &  Q  &  Y \\
N& 10 18 08.75 & 50 05 47.4 & 5.9 & 21.25 &  2.87 & -0.05 & -0.6 &   310 & (52.9) &(0.79)&        &       29.2  &EC1(c)& C \\
 & 10 18 11.08 & 44 50 08.4 & 1.8 & 20.73 &  1.19 & -0.01 & -0.3 &   240 &   54   & 0.65 &   54.0 &  $\leq0.3$  &  Q  &  N \\
 & 10 18 24.11 & 34 52 29.2 & 2.3 & 17.77 &   .56 & -0.07 & -0.5 &  1020 &(457.4) &(0.36)&  371.0 &       19.0  &  Q  &  M \\
 & 10 18 41.01 & 40 46 55.9 & 2.5 & 19.27 &   .99 &  0.00 &  1.2 &   860 &  128   & 0.83 &    3.7 &       11.0  & QC  &  Y \\
 & 10 18 47.80 & 32 42 46.2 & 1.7 & 20.13 &  1.16 &  0.04 & -0.4 &  2500 &  247   & 1.01 &   54.0 &        9.0  &  Q  &  Y \\
N& 10 19 05.59 & 46 11 20.1 & 1.6 & 21.51 &  2.00 & -0.23 & -0.8 &   860 & (129.6)&(0.85)&        &      (20.9) &EC1(c)  &  C \\
 & 10 19 12.25 & 37 13 47   & 6.2 & 20.82 &  1.06 &  0.10 &  4.6 &  1320 &  233   & 0.76 &        &       30.0  &  M  &  Y \\
 & 10 19 31.30 & 37 50 45.0 & 8.8 & 20.90 &  1.34 & -0.01 &  2.5 &   250 &   14   & 1.27 &   10.1 &        2.0  & QC  &  y \\
 & 10 19 33.07 & 45 56 15.4 & 8.0 & 20.71 &  1.18 &  0.04 & -0.6 &   650 &  134   & 0.69 &   23.0 &       42.0  &  Q  &  Y \\
 & 10 19 40.56 & 39 47 00.3 & 9.3 & 17.04 &   .79 &  0.10 & -0.4 &   310 &  115   & 0.43 &   42.0 &       25.0  &  Q  &  Y \\
 & 10 19 56.13 & 34 06 08.5 & 1.6 & 19.53 &  1.00 & -0.05 & -0.3 &   670 &   80   & 0.93 &    2.3 &       37.0  &  Q  &  Y \\
N& 10 20 03.24 & 43 40 24.0 & 3.8 & 21.68 &  3.78 &  0.05 &  3.3 &   430 & (109.4)&(0.61)&        &     ($<17$) &EC1  &  C \\
\hline\hline
\end{tabular}
{\caption[Table 1]{\label{tab:tab1} {\bf (cont).}
}}
\end{center}
\end{table*}
\normalsize

\addtocounter{table}{-1}
\clearpage

\scriptsize
\begin{table*}
\begin{center}
\begin{tabular}{|l|ll|r|r|r|r|r|r|r|r|r|r|r|l}
\hline\hline
(1) & (2) & (3) & (4) & (5) & (6) & (7) & (8) & (9) & (10) & (11) & (12) & (13) & (14) & (15) \\
\hline
& \mc{1}{c}{RA(1950)} & \mc{1}{c|}{Dec(1950)} & \mc{1}{c|}{$r$}
& \mc{1}{c|}{$O$} & \mc{1}{c|}{$O-E$} & \mc{1}{c|}{$\Delta \rm RA$} & \mc{1}{c|}{$\Delta \rm Dec$} 
& \mc{1}{c|}{$S_{151}$} & \mc{1}{c|}{$S_{1490}$} & \mc{1}{c|}{$\alpha_{151}^{1490}$} &
\mc{1}{c|}{$S_{\rm core}$} & \mc{1}{c|}{$\theta$} & \mc{1}{c|}{Opt} & 
\\
\hline
 & 10 20 05.94 & 48 06 57.5 & 1.6 & 19.70 &   .82 & -0.04 &  1.1 &  2570 &  470   & 0.74 &  470.0 &        0.6  &Q/EC1&  Y \\
 & 10 20 14.6  & 40 03 28.0 & 4.0 & 17.38 &   .70 &  0.11 &  1.7 &  2960 & (1123) &(0.44)&        &       $<5$  &  Q  &  V \\
R& 10 20 14.6  & 40 03 28.0 & 4.0 & 17.25 &   .54 & -0.07 &  1.8 &  2960 & (1123) &(0.44)&        &       $<5$  &  Q  &  V \\
 & 10 20 17.31 & 43 47 15.8 & 9.0 & 12.37 &   .98 & 1.19  &  4.2 &   280 & (33.0) &(0.96)&        &     ($<19$) &  S  &  C \\
 & 10 20 21.70 & 36 19 46.9 & 2.5 & 20.18 &  1.01 & -0.05 &  0.3 &   140 & (29.2) &(0.70)&   19.1 &       16.0  &  Q  &  F \\
 & 10 20 24.68 & 48 39 48.7 & 0.4 & 20.32 &   .53 &  0.02 & -0.1 &  2200 & (240)  &(0.99)&    5.0 &       43.0  &  Q  &  Y \\
 & 10 21 14.24 & 39 53 13.9 & 8.0 & 20.13 &  1.14 & -0.10 & 17.9 &   130 &   17   & 0.89 &        &        2.0  &  M  &  Y \\
R& 10 21 14.24 & 39 53 13.9 & 9.4 & 20.30 &  1.21 & -0.29 & 18.0 &   130 &   17   & 0.89 &        &        2.0  &  M  &  Y \\
N& 10 21 57.99 & 37 08 33.0 & 1.0 & 20.84 &  1.30 &  0.50 & 11.4 &    90 & (19.3) &(0.69)&        &      (27.9) &  ?  &  C$\ddag$\\
 & 10 22 07.62 & 39 19 04.8 & 5.2 & 21.39 &  1.41 & -0.03 & -0.3 &   120 &   24   & 0.70 &   24.4 &        1.2  &  Q  &  Y \\
 & 10 22 46.73 & 43 32 51.5 & 6.9 & 18.40 &  1.09 & -0.07 &  1.2 &   430 & (57.5) &(0.90)&    5.0 &       35.0  &  QC &  y \\
N& 10 22 52.10 & 35 06 06.2 & 5.0 & 20.37 &  2.69 & -0.10 & -0.4 &   550 & (110)  &(0.72)&        &     (28.9)  &EC1(c)  &  F \\
 & 10 23 08.98 & 35 53 52.3 & 2.2 & 19.44 &  1.24 &  0.41 &-20.8 &   130 & (20.2) &(0.84)&    8.5 &        0.7  &  M  &  N \\
 & 10 23 51.31 & 37 13 43.1 & 1.4 & 20.98 &  1.77 & -0.05 &  0.0 &   180 &  173   & 0.02 &  173.0 &  $\leq0.3$  &  Q  &  N \\
 & 10 24 01.64 & 41 54 42.9 & 7.6 & 18.63 &   .45 & -0.04 &  0.1 &   490 & (90.1) &(0.76)&    1.1 &       50.0  &  Q  &  y \\
 & 10 24 06.92 & 46 46 14.3 & 9.1 & 19.81 &   .42 &  0.04 & -0.5 &   130 &   16   & 0.92 &   14.4 &        2.2  & QC  &  Y \\
 & 10 24 09.47 & 48 18 31.7 & 0.6 & 19.46 &   .67 &  0.02 &  0.3 &   770 &  233   & 0.52 &  164.0 &        6.0  &  Q  &  Y \\
N& 10 24 14.39 & 39 23 24.6 & 1.8 & 21.22 &  4.84 & -0.16 &  1.9 &   420 & (43.9) &(1.01)&        &     ($<19$) &EC2  &  C \\
 & 10 24 26.59 & 43 07 19.7 & 5.3 & 19.23 &   .68 & -0.07 &  0.8 &   300 &   59   & 0.71 &        &       10.0  & QC  &  Y \\
N& 10 24 29.6  & 48 32 46   & 8.9 & 20.98 &  3.24 & -0.22 &  8.5 &  6600 & (987)  &(0.85)&        &     (69.0)  &EC1  &  C \\
 & 10 24 54.03 & 36 12 51.7 & 2.5 & 18.97 &   .83 & -0.01 & -0.3 &   200 & (47.1) &(0.65)&    5.2 &       61.0  & QC  &  Y \\
 & 10 25 34.59 & 43 21 47.5 & 3.3 & 19.20 &   .82 & -0.08 &  1.0 &   550 &   94   & 0.77 &   10.3 &       39.0  &  Q  &  Y \\
N& 10 26 36.88 & 37 49 08.8 & 9.4 & 18.84 &  2.22 & -0.73 & 13.9 &   250 & (66.1) &(0.60)&        &   $\leq 4$  &  M  &  F \\
 & 10 27 08.83 & 32 39 42.8 & 9.7 & 19.71 &   .57 &  0.00 & -0.6 &   210 &  100   & 0.32 &   98.0 &        4.0  &  Q  &  Y \\
 & 10 27 23.44 & 43 24 31.7 & 2.8 & 19.72 &   .72 & -0.03 &  0.5 &   700 &  115   & 0.79 &   36.0 &        8.0  &  Q  &  Y \\
 & 10 27 24.64 & 34 01 56.0 & 4.2 & 20.41 &   .59 &  0.02 & -0.9 &   160 &   14   & 1.06 &        &        2.0  &Q/EC1&  Y \\
 & 10 28 10.62 & 41 34 00.9 & 9.7 & 20.97 &  2.90 &  0.64 &  6.3 &   130 &   25   & 0.71 &   25.2 &        0.4  &  M  &  y \\
 & 10 28 47.81 & 42 09 46.9 & 3.1 & 18.21 &   .69 & -0.03 &  0.3 &  1610 &  215   & 0.88 &    2.1 &       41.0  &  Q  &  Y \\
N& 10 28 59.3  & 35 17 44   & 9.2 & 21.38 &  1.83 & -0.66 &  4.6 &  3000 &   (c)  &      &        &       7C(c) &  M  &  D \\
 & 10 29 31.32 & 49 48 11.0 & 0.9 & 21.06 &  1.28 &  0.07 &  1.0 &   200 &   46   & 0.64 &    9.0 &       16.0  &  Q  &  y \\
 & 10 29 47.42 & 37 53 54.4 & 1.7 & 18.04 &   .86 &  0.05 &  0.8 &   320 &   72   & 0.65 &   37.0 &        6.0  & QC  &  Y \\
 & 10 29 52.44 & 39 25 37.8 & 7.0 & 18.77 &   .66 & -0.06 & -0.3 &   120 & (25.2) &(0.70)&    6.4 &        6.6  &  Q  &  Y \\
 & 10 30 07.80 & 41 31 34.5 & 4.0 & 18.36 &  1.00 & -0.05 &  0.4 &   980 &  547   & 0.25 &  481.0 &        5.0  &  Q  &  Y \\
 & 10 31 09.6  & 49 46 40   & 8.2 & 21.02 &  1.25 & -0.43 & -0.9 &   160 &   31   & 0.72 &        &        8.0  &  M  &  Y \\
 & 10 32 21.14 & 45 26 17.9 & 5.5 & 19.22 &  1.85 &  0.15 & 10.4 &   160 &   26   & 0.79 &        &        7.3  &  M  &  Y \\
 & 10 32 21.21 & 34 21 56.9 & 2.1 & 18.34 &   .41 &  0.06 & -1.3 &   510 &  117   & 0.64 &   44.0 &       44.0  & QC  &  Y \\
 & 10 32 34.50 & 44 25 03.7 & 5.8 & 20.37 &  1.66 & -0.07 &  0.7 &   120 &   36   & 0.52 &   24.9 &        5.0  & QC  &  Y \\
 & 10 32 58.36 & 38 12 14.5 & 7.9 & 17.14 &   .26 &  0.10 &  1.3 &   140 &   54   & 0.42 &   45.0 &        2.0  &  Q  &  Y \\
 & 10 33 13.62 & 48 22 44.4 & 8.0 & 21.21 &  1.25 & -0.07 &  8.4 &   440 &   73   & 0.78 &   73.0 &        1.3  &  M  &  Y \\
N& 10 33 25.15 & 36 49 54.3 & 8.2 & 20.88 &  2.31 &  0.05 &  0.3 &    90 & (75.5) &(0.08)&        &     ($<18$) &EC1(c)& C$\ddag$\\
 & 10 33 50.91 & 33 36 41.3 & 7.4 & 19.15 &  1.08 &  0.16 &  9.5 &   100 & (23.2) &(0.66)&        &     ($<20$) &OC1  &  C$\ddag$\\
N& 10 33 59.50 & 34 27 00.2 & 8.8 & 21.55 &  1.72 & -0.87 & 15.6 &   130 & (18.6) &(0.87)&        &     (27.4)  &  M  &  F \\
N& 10 34 00.75 & 38 47 13.2 & 9.6 & 20.67 &  2.93 &  0.11 &  5.1 &   130 & (19.6) &(0.85)&        &     (28.8)  &OC1  &  F \\
 & 10 34 08.5  & 46 09 23   & 7.8 & 20.70 &  2.83 & -0.63 & 45.8 &   230 &   18   & 0.84 &        &        6.0  &  M  &  y \\
T& 10 34 10.1  & 46 07 46   & 7.8 &       &       &       &      &       &    3   &      &        &        2.0  &  M  &  y \\
T& 10 34 17.2  & 46 09 58   & 7.8 &       &       &       &      &       &   12   &      &        &        5.0  &  M  &  y \\   
 & 10 34 15.93 & 47 05 49.4 & 0.5 & 19.88 &   .56 &  0.01 & -0.1 &   500 &   89   & 0.75 &   52.0 &        4.0  &Q/OC1&  Y \\
R& 10 34 15.93 & 47 05 49.4 & 0.5 & 19.17 &   .52 &  0.00 &  0.2 &   500 &   89   & 0.75 &   52.0 &        4.0  &Q/OC1&  Y \\
N& 10 34 22.63 & 43 50 51.0 & 8.7 & 11.86 &  3.28 & -0.25 &  0.9 &   210 &(132.5) &(0.21)&        &      ($<19$)&  G  &  C \\
N& 10 34 22.42 & 45 40 14.3 & 9.9 & 21.39 &  1.53 & -1.11 &-12.5 &    90 & (11.8) &(0.91)&        &      ($<31$)&EC1  &  C$\ddag$\\
R& 10 34 22.42 & 45 40 14.3 & 9.5 & 20.55 &  1.03 & -1.32 &-10.8 &    90 & (11.8) &(0.91)&        &      ($<31$)&  ?  &  C$\ddag$\\
N& 10 34 39.52 & 39 43 28.5 & 8.8 & 19.40 &  2.64 & -1.45 & -0.1 &   230 & (28.7) &(0.93)&        &         42  &  M  &  F\\
N& 10 34 50.86 & 45 44 07.2 & 5.4 & 21.76 &  1.81 &  0.28 & -4.1 &    90 & (21.3) &(0.65)&        &      (19.7) &OC1(c)& C$\ddag$\\
 & 10 35 06.94 & 38 56 12.9 & 2.8 & 17.50 &   .64 & -0.22 & 17.6 &   170 &   18   & 0.98 &        &        2.3  &  M  &  Y \\
 & 10 35 17.4  & 48 41 24   & 2.2 & 19.77 &   .84 & -0.05 &  0.8 &  2080 &  226   & 0.97 &        &       22.0  &  Q  &  Y \\
 & 10 35 36.99 & 50 03 13.7 & 1.5 & 20.93 &  1.03 &  0.04 &  0.5 &  1070 &  124   & 0.94 &    4.9 &       24.0  &  Q  &  Y \\
R& 10 35 36.99 & 50 03 13.7 & 2.6 & 20.93 &  1.80 & -0.13 &  2.3 &  1070 &  124   & 0.94 &    4.9 &       24.0  &  Q  &  Y \\
N& 10 35 47.50 & 46 33 17.9 & 4.9 & 21.09 &  1.35 & -0.11 &  1.8 &   150 & (36.2) &(0.64)&        &      (16.6) &EC1(c)&  C \\
\hline\hline
\end{tabular}
{\caption[Table 1]{\label{tab:tab1} {\bf (cont).}
}}
\end{center}
\end{table*}
\normalsize

\addtocounter{table}{-1}
\clearpage

\scriptsize
\begin{table*}
\begin{center}
\begin{tabular}{|l|ll|r|r|r|r|r|r|r|r|r|r|r|l}
\hline\hline
(1) & (2) & (3) & (4) & (5) & (6) & (7) & (8) & (9) & (10) & (11) & (12) & (13) & (14) & (15) \\
\hline
& \mc{1}{c}{RA(1950)} & \mc{1}{c|}{Dec(1950)} & \mc{1}{c|}{$r$}
& \mc{1}{c|}{$O$} & \mc{1}{c|}{$O-E$} & \mc{1}{c|}{$\Delta \rm RA$} & \mc{1}{c|}{$\Delta \rm Dec$} 
& \mc{1}{c|}{$S_{151}$} & \mc{1}{c|}{$S_{1490}$} & \mc{1}{c|}{$\alpha_{151}^{1490}$} &
\mc{1}{c|}{$S_{\rm core}$} & \mc{1}{c|}{$\theta$} & \mc{1}{c|}{Opt} & 
\\
\hline
 & 10 36 00.86 & 49 47 16.1 & 0.4 & 21.11 &  2.24 &  0.01 &  1.7 &   210 & (34.7) &(0.81)&        &      (37.5) &  ?(c)&  C$^{r}$ \\
 & 10 36 16.57 & 47 21 18.9 & 1.2 & 20.98 &  2.08 & -0.04 & -0.1 &  3800 &(381.6) &(1.03)&        &     ($<19$) &  ?(c)&  C$^{r}$ \\
 & 10 36 23.04 & 41 19 42.9 & 4.8 & 21.24 &  2.96 &  0.06 & -1.3 &   160 &   22   & 0.86 &   22.4 &        0.7  & QC  &  N \\
 & 10 36 36.52 & 50 07 18.3 & 7.6 & 19.83 &  1.16 &  0.61 & -1.4 &   290 & (33.5) &(0.97)&        &     ($<20$) & QC  &  C \\
N& 10 36 44.43 & 48 18 22.9 & 7.3 & 20.87 &  3.29 & -0.93 &  7.6 &   240 & (22.6) &(1.06)&        &     ($<26$) &  ?  &  C$^{r}$ \\
 & 10 37 01.66 & 42 41 17.4 & 0.5 & 20.38 &  1.33 & -0.11 &  0.9 &   700 &  107   & 0.82 &    2.2 &       25.0  &Q/EC1&  Y \\
 & 10 37 25.93 & 45 05 16.7 & 6.5 & 19.11 &   .52 & -0.09 & -0.5 &   110 & (23.7) &(0.69)&   10.0 &        1.0  &Q/OC1&  Y \\
 & 10 37 44.72 & 45 45 13.2 & 7.8 & 18.51 &   .70 & -0.06 &  0.4 &   140 & (60.0) &(0.38)&        &     ($<19$) &  Q  &  C$^{j}$ \\
 & 10 38 09.41 & 43 13 05.1 & 1.2 & 19.12 &   .79 & -0.01 &  0.6 &   790 &  100   & 0.90 &    8.0 &       10.0  &  Q  &  Y \\
N& 10 38 47.48 & 40 05 01.0 & 4.9 & 21.24 &  3.27 &  0.06 & -0.5 &   100 & (35.8) &(0.46)&    3.8 &       38.0  &EC1(c)&F$\ddag$\\
 & 10 38 51.52 & 46 49 16.9 & 5.5 & 19.37 &   .40 & -0.03 & -1.1 &   120 &   29   & 0.63 &   28.6 &        1.1  &  Q  &  Y \\
N& 10 39 01.49 & 46 31 56.1 & 4.4 & 20.65 &  2.61 &  0.03 &  0.9 &   120 & (28.7) &(0.64)&        &     ($<22$) &EC1(c)& C \\
 & 10 41 40.55 & 49 57 27.2 & 4.6 & 19.62 &   .58 & -0.08 &  0.9 &   820 &  116   & 0.85 &   22.0 &       26.0  &  Q  &  Y \\
 & 10 42 09.71 & 45 49 47.8 & 8.5 & 19.93 &  1.16 &  0.28 & -0.7 &   710 & (91.3) &(0.92)&        &     ($<17$) & QC  &  C \\
N& 10 42 09.91 & 41 30 16.5 & 2.4 & 21.52 &  2.77 &  0.60 &  6.7 &   130 & (14.5) &(0.98)&        &       $<5$  &EC1  &  F \\
 & 10 42 22.76 & 49 03 53.0 & 3.4 & 20.02 &   .24 & -0.08 &  0.7 &   200 &   26   & 0.89 &        &        2.0  &  Q  &  Y \\
N& 10 42 22.36 & 39 16 20.8 & 9.6 & 20.25 &  2.49 & -0.70 & 17.2 &   380 & (50.5) &(0.91)&        &       $<6$  &  M  &  F \\
N& 10 42 28.99 & 39 20 21.2 & 9.9 & 20.81 &  2.08 & -0.45 & 25.7 &   540 & (76.3) &(0.88)&        &       55.0  &  M  &  F \\
 & 10 42 53.0  & 41 24 57   & 1.9 & 18.54 &   .60 &  0.02 &  1.1 &  1580 &  245   & 0.81 &        &        8.5  &  Q  &  Y \\
 & 10 42 57.57 & 39 18 54.3 & 0.2 & 19.85 &   .24 &  0.05 & -0.7 &   110 & (14.8) &(0.90)&    4.5 &        4.0? &  Q  &  y \\
N& 10 43 00.78 & 48 32 34.7 & 5.8 & 18.52 &  1.85 &  0.51 &  4.2 &   100 & (19.3) &(0.74)&        &     ($<27$) &EC2  &  C$\ddag$\\
 & 10 43 12.7  & 42 38 18   & 2.0 & 20.57 &  1.68 &  0.10 & -0.4 &   350 &   66   & 0.73 &        &        3.1  &  Q  &  Y \\
 & 10 43 20.58 & 45 27 54.1 & 6.5 & 18.67 &   .73 & -0.05 & -0.1 &   360 &   42   & 0.94 &   22.4 &       19.0  & QC  &  Y \\
 & 10 43 59.23 & 48 43 13.9 & 3.2 & 20.45 &  2.34 & -0.06 &  1.7 &   100 & (38.9) &(0.42)&        &     ($<21$) &  G  & C$\ddag$\\
 & 10 44 13.24 & 44 32 32.1 & 0.9 & 19.68 &   .87 &  0.02 & -0.1 &   350 &   57   & 0.79 &    5.1 &       29.0  &G/OC2&  Y \\
R& 10 44 13.24 & 44 32 32.1 & 0.8 & 19.77 &   .83 & -0.04 &  0.2 &   350 &   57   & 0.79 &    5.1 &       29.0  &  Q  &  Y \\
 & 10 44 22.08 & 41 46 00.8 & 5.4 & 18.67 &   .61 &  0.03 & -0.5 &   140 &   38   & 0.57 &   29.0 &       14.0  & QC  &  Y \\
 & 10 44 34.38 & 48 51 21.1 & 4.0 & 19.88 &   .77 & -0.10 &  1.0 &   170 &   58   & 0.47 &   43.4 &        9.4  &Q/OC1&  Y \\
 & 10 44 35.76 & 47 41 22.4 & 1.7 & 18.48 &   .88 & -0.10 &  1.1 &  2000 & (789)  &(0.42)&        &        0.6  &  Q  &  P \\
S& 10 44 35.76 & 47 41 22.4 & 9.8 & 20.30 &  1.77 & -0.43 &  9.2 &  2000 & (789)  &(0.42)&        &        0.6  &  M  &  P \\ 
 & 10 45 08.13 & 48 18 12.7 & 9.6 & 20.35 &  1.53 & -0.14 &  0.7 &    90 & (19.4) &(0.69)&        &     ($<25$) &EC2(c) & C$\ddag$\\ 
 & 10 45 09.35 & 49 55 19.8 & 7.7 & 19.76 &   .52 & -0.09 & -1.9 &   100 & (24.3) &(0.64)&        &      (20.6) &  ?(c) & C$\ddag$\\
S& 10 45 09.35 & 49 55 19.8 & 7.8 & 21.00 &  3.55 &  1.20 & -6.4 &   100 & (24.3) &(0.64)&        &      (20.6) &  M    & C$\ddag$\\
N& 10 45 11.75 & 48 35 42.5 & 2.6 & 10.83 &   .90 & -0.18 & -1.0 &    90 & (12.0) &(0.90)&        &     ($<31$) &  ?(c) & C$\ddag$\\ 
 & 10 45 16.88 & 38 53 26.8 & 8.1 & 17.58 &   .35 &  0.08 & -1.2 &   200 &   35   & 0.76 &    5.7 &       18.0  &  Q    &  Y \\
N& 10 45 30.78 & 42 16 56.5 & 9.1 & 20.23 &  2.14 & -0.10 & -9.5 &   450 & (40.8) &(1.08)&        &     (36.0)  &EC1    &  C \\
N& 10 45 59.60 & 46 43 56.7 & 6.2 & 21.18 &  3.11 & -0.01 & -2.9 &   310 & (63.4) &(0.71)&        &     (31.5)  &EC2(c) & C \\
 & 10 46 00.47 & 46 29 09.6 & 3.0 & 20.68 &  3.08 & -0.21 & -5.7 &  1290 &(128.3) &(1.04)&        &     (40.9)  &S      &  C \\
N& 10 46 14.00 & 44 56 47.8 & 7.5 & 21.40 &  2.77 &  0.35 & 15.3 &   150 & (22.3) &(0.86)&        &     (30.4)  &EC1    &  C \\
N& 10 46 22.32 & 46 27 26.8 & 5.1 & 19.72 &  2.29 & -0.51 & -5.5 &   220 & (21.9) &(1.04)&        &    ($<20$)  &EC1    &  C \\
 & 10 46 31.93 & 46 11 42.6 & 1.8 & 18.52 &   .47 & -0.08 & -0.6 &   500 &   63   & 0.90 &    5.1 &       29.0  &  Q    &  Y \\
N& 10 46 40.30 & 45 01 36.3 & 5.5 & 21.07 &  2.59 &  0.60 &  1.1 &   300 & (63.6) &(0.70)&        &     (51.6)  &EC1    &  C \\
 & 10 46 55.3  & 50 09 07   & 0.5 & 20.08 &  1.03 &  0.17 & -7.6 &   290 &   40   & 0.87 &        &       67.0  & QC    &  Y \\
N& 10 48 20.64 & 47 00 12.7 & 1.3 & 18.82 &  1.99 & -0.18 & -0.4 &   930 &(321.0) &(0.48)&        &        1.5  &EC1(c)& P \\
N& 10 49 04.97 & 39 21 00.1 & 9.0 & 20.47 &  3.01 &  0.29 &-11.3 &    80 & (12.4) &(0.81)&        &     (67.3)  &OC1  &  C$\ddag$\\
 & 10 49 30.41 & 47 50 16.9 & 6.2 & 20.80 &   .81 & -0.04 &  0.6 &   220 &   41   & 0.73 &   25.8 &        5.0  & QC  &  Y \\
 & 10 49 34.13 & 47 37 16.9 & 0.7 & 19.60 &   .94 & -0.07 &  0.7 &   140 &   23   & 0.78 &    3.0 &       20.0  &  Q  &  Y \\
N& 10 49 39.37 & 45 40 14.8 & 7.4 & 21.35 &  1.61 &  0.78 & 22.8 &   100 & (26.3) &(0.60)&        &     (41.8)  &  ?  &  C$\ddag$\\
 & 10 49 39.78 & 48 55 58.1 & 7.7 & 17.83 &   .64 &  0.74 & -4.2 &   120 &   23   & 0.73 &    2.2 &       33.0  &  M  &  y \\
 & 10 50 01.46 & 46 03 46.7 & 5.3 & 18.68 &   .68 &  0.44 & -6.2 &   150 & (23.6) &(0.83)&        &     (22.0)  &Q/EC1&  n \\
 & 10 50 07.81 & 43 18 52.4 & 6.5 & 18.50 &  1.17 & -1.25 & -6.7 &   170 &   37   & 0.67 &    4.1 &       34.0  &  G  &  y \\
N& 10 51 01.42 & 42 47 12.1 & 3.2 & 21.14 &  2.25 &  0.24 &  2.7 &   160 & (21.4) &(0.90)&        &     (39.9)  &OC2(c)& C \\
 & 10 51 15.45 & 47 30 39.8 & 5.8 & 21.07 &  2.99 &  0.34 & -0.4 &   120 & (22.5) &(0.75)&        &     (24.6)  &EC2(c)& C \\
S& 10 51 15.45 & 47 30 39.8 & 7.6 & 20.49 &  2.48 &  0.11 &  9.2 &   120 & (22.5) &(0.75)&        &     (24.6)  &  M  &  C \\
 & 10 51 38.56 & 45 57 51.9 & 3.8 & 19.19  &  .37 & -0.01 &  0.1 &   350 &   49   & 0.86 &   49.0 &        1.5  &  Q  &  Y \\
 & 10 52 17.03 & 46 32 17.0 & 4.7 & 20.03 &  2.31 &  0.45 & 13.5 &   130 & (18.0) &(0.89)&        &     (46.8)  &OC1  &  C \\
N& 10 52 17.13 & 46 36 53.9 & 7.8 & 21.20 &  2.71 & -3.36 &-13.3 &   840 & (49.4) &(1.27)&        &     (35.8)  &  ?  &  C$^{j}$ \\
N& 10 53 00.43 & 47 15 47.6 & 8.1 & 20.40 &  2.53 & -0.02 &  0.4 &   120 & (28.2) &(0.65)&        &     (35.6)  &EC1(c)& C \\

\hline\hline
\end{tabular}
{\caption[Table 1]{\label{tab:tab1} {\bf (cont).}
}}
\end{center}
\end{table*}
\normalsize

\addtocounter{table}{-1}
\clearpage

\scriptsize
\begin{table*}
\begin{center}
\begin{tabular}{|l|ll|r|r|r|r|r|r|r|r|r|r|r|l}
\hline\hline
(1) & (2) & (3) & (4) & (5) & (6) & (7) & (8) & (9) & (10) & (11) & (12) & (13) & (14) & (15) \\
\hline
& \mc{1}{c}{RA(1950)} & \mc{1}{c|}{Dec(1950)} & \mc{1}{c|}{$r$}
& \mc{1}{c|}{$O$} & \mc{1}{c|}{$O-E$} & \mc{1}{c|}{$\Delta \rm RA$} & \mc{1}{c|}{$\Delta \rm Dec$} 
& \mc{1}{c|}{$S_{151}$} & \mc{1}{c|}{$S_{1490}$} & \mc{1}{c|}{$\alpha_{151}^{1490}$} &
\mc{1}{c|}{$S_{\rm core}$} & \mc{1}{c|}{$\theta$} & \mc{1}{c|}{Opt} & 
\\
\hline
*& 10 53 18.00 & 44 43 18.1 &20.1 & 19.51 &   .30 & -0.08 &  0.8 &   370 &  100   & 0.57 &   55.0 &       29.0  &OC2(c)& y \\    
S& 10 53 18.00 & 44 43 18.1 & 7.0 & 19.66 &   .89 & -2.00 &  4.5 &   370 &  100   & 0.57 &   55.0 &       29.0  &  M  &  y \\
S& 10 53 18.00 & 44 43 18.1 & 5.5 & 19.70 &   .71 & -2.03 & 13.7 &   370 &  100   & 0.57 &   55.0 &       29.0  &  M? &  y \\
N& 10 53 29.14 & 48 42 06.3 & 8.1 & 21.46 &  2.80 &  1.00 & 10.8 &   120 & (28.6) &(0.64)&        &     (23.6)  &EC1  &  C \\
 & 10 53 33.44 & 49 36 43.0 & 9.1 & 20.30 &   .97 &  0.00 &  1.8 &   110 & (12.8) &(0.97)&        &    ($<32$)  &?(c) &  C$^{j}$ \\ 
 & 10 53 52.3  & 49 49 43   & 6.3 & 20.94 &  1.54 &  0.05 &  2.2 &  1030 &  130   & 0.90 &        &       24.0  & QC  &  Y \\
 & 10 54 25.55 & 46 12 18.2 & 4.2 & 19.59 &   .52 &  0.00 & -0.5 &   130 &   13   & 1.01 &    2.0 &       17.0  &Q/OC2&  y \\
 & 10 55 17.75 & 49 55 40.0 & 2.4 & 18.73 &   .76 & -0.07 &  1.3 &  1250 &  243   & 0.71 &        &        3.0  &  Q  &  L \\
N& 10 56 51.38 & 48 25 07.4 & 2.9 & 20.78 &  3.14 &  0.26 &  4.5 &   100 & (26.6) &(0.59)&        &     (20.5)  &EC1(c)& C$\ddag$\\
N& 10 57 25.22 & 46 27 08.0 & 0.7 & 20.05 &  1.23 & -0.07 &  1.5 &   670 &(177.3) &(0.60)&        &    ($<18$)  &EC1  &  C \\
 & 10 58 32.56 & 49 06 38.6 & 2.7 & 20.66 &   .98 & -0.03 &  1.4 &   250 &   40   & 0.80 &    7.7 &       23.0  &  Q  &  Y \\
N& 10 59 15.24 & 47 59 36.1 & 9.5 & 18.10 &  2.11 &  1.02 &-13.2 &   100 & (10.4) &(1.02)&        &    ($<37$)  &  ?  &  C$\ddag$\\
 & 10 59 58.3  & 48 20 02   & 1.8 & 20.31 &   .97 & -0.11 &  0.7 &   210 &   42   & 0.70 &        &       28.0  & QC  &  Y \\
 & 11 00 37.65 & 46 30 10.9 & 6.1 & 18.97 &   .72 & -0.17 & -1.6 &   180 & (52.6) &(0.55)&        &     (20.7)  &  Q  &  C$^{j}$ \\
 & 11 00 48.33 & 46 06 19.4 & 7.9 & 20.20 &  1.19 &  0.62 & -6.9 &   140 & (13.9) &(1.04)&        &     (43.4)  &  ?  &  C$^{l}$ \\
N& 11 01 25.07 & 44 59 20.5 & 3.9 & 21.35 &  3.70 & -0.08 & -1.0 &   110 & (14.7) &(0.90)&        &    ($<29$)  &EC1(c)& C \\
 & 11 01 33.79 & 49 44 35.5 & 2.1 & 20.06 &  1.34 & -0.09 &  2.2 &   870 &  211   & 0.62 &  211.0 &        0.1  & QC  &  L \\
N& 11 01 52.03 & 47 43 05.8 & 6.5 & 20.47 &  3.55 & -0.82 & -3.2 &   400 & (87.4) &(0.68)&        &     (44.3)  &EC1  &  C \\
 & 11 02 50.23 & 46 35 40.8 & 2.3 & 20.93 &  2.35 & -0.02 &  0.1 &   220 & (40.2) &(0.76)&        &    ($<20$)  & ?(c)&  C$^{r}$\\  
N& 11 07 53.72 & 48 03 22.8 & 2.3 & 19.10 &  2.47 & -0.15 & -1.3 &   880 &(122.6) &(0.89)&        &     (67.8)  &EC1(c)& C \\

\hline\hline
\end{tabular}
{\caption[Table 1]{\label{tab:tab1} {\bf (cont).}
}}
\end{center}
\end{table*}
\normalsize

\scriptsize
\begin{table*}
\begin{center}
\begin{tabular}{|l|ll|r|r|r|r|r|r|r|r|r|r|r|l}
\hline\hline
(1) & (2) & (3) & (4) & (5) & (6) & (7) & (8) & (9) & (10) & (11) & (12) & (13) & (14) & (15) \\
\hline
& \mc{1}{c}{RA(1950)} & \mc{1}{c|}{Dec(1950)} & \mc{1}{c|}{$r$}
& \mc{1}{c|}{$O$} & \mc{1}{c|}{$O-E$} & \mc{1}{c|}{$\Delta \rm RA$} & \mc{1}{c|}{$\Delta \rm Dec$} 
& \mc{1}{c|}{$S_{151}$} & \mc{1}{c|}{$S_{1490}$} & \mc{1}{c|}{$\alpha_{151}^{1490}$} &
\mc{1}{c|}{$S_{\rm core}$} & \mc{1}{c|}{$\theta$} & \mc{1}{c|}{Opt} & 
\\
\hline
 & 09 53 27.30 & 41 00 44.7 & 12.4 & 21.06 &  1.51 & -0.06 & -0.2 &  260  &  66 &  0.60 &    35.0 &       11.0  & QC  &  Y \\
 & 09 54 55.71 & 44 20 23.7 & 10.4 & 19.47 &  0.56 & -0.10 &  2.0 & 1480  & 212 &  0.85 &     7.0 &       16.0  &  Q  &  y \\
 & 09 55 33.1  & 42 32 29   &  5.8 & 18.46 &  1.26 &  0.65 & -6.6 &  520  &(72.2)&(0.89)&         &       39.0  &  M  &  y \\
 & 10 00 00.23 & 41 02 44.0 & 10.8 & 20.17 &  0.88 & -0.05 &  0.4 &  130  &  23 &  0.76 &     8.0 &       24.0  &  Q  &  Y \\
 & 10 03 51.64 & 45 10 16.1 & 10.2 & 17.35 &  0.10 & -0.66 &  3.8 & 1500  & 222 &  0.83 &    77.0 &        7.7  &  M  &  Y \\
 & 10 04 53.6  & 48 40 04   &  9.9 & 20.24 &  1.35 &  1.05 &  5.5 &  300  &  42 &  0.86 &         &       11.3  &  M  &  y \\
 & 10 05 47.68 & 46 27 44.2 & 14.6 & 15.37 &  0.75 &  1.59 & 24.1 &  880  & 144 &  0.79 &     2.0 &       54.0  &  M  &  Y \\
 & 10 07 26.09 & 41 47 25.5 & 13.5 & 16.21 &  0.24 & -0.02 &  0.7 & 8300  &(1236)&(0.86)&   153.0 &       33.0  &  Q  &  S \\
 & 10 09 21.46 & 40 44 24.2 & 10.3 & 19.56 &  1.36 & -0.67 & -2.0 &  130  &  13 &  1.01 &         &        3.5  &  M  &  y \\
 & 10 10 24.61 & 36 10 25.1 & 11.1 & 20.72 &  1.63 & -1.36 & -8.2 &  120  &  10 &  1.09 &    10.0 &        0.5  &  M  &  N \\
 & 10 13 01.12 & 47 23 10.2 &  7.8 & 19.63 &  3.18 & -0.07 &  0.0 &  300  &  41 &  0.87 &     1.4 &       20.0  &  G  &  y \\
 & 10 15 28.8  & 38 20 32   & 10.4 & 17.93 &  0.21 & -0.09 & -3.3 & 1200  &(274)& (0.66)&         &       43.0  &  Q  &  V \\
 & 10 15 52.81 & 48 23 37.1 & 11.5 & 20.52 &  1.15 & -0.63 & -6.5 &  120  &  18 &  0.83 &    18.0 &        0.9  &  M  &  Y \\
 & 10 17 08.68 & 43 38 51.8 & 13.8 & 20.01 &  1.11 & -0.57 & -4.4 &  120  &  26 &  0.67 &    26.0 &        0.6  &  M  &  N \\
 & 10 19 03.37 & 41 41 14.4 &  9.5 & 14.85 &  0.09 &  0.24 &-10.2 &  430  & 195 &  0.35 &   154.0 &        4.5  &  M  &  y \\
 & 10 20 28.96 & 44 29 26.0 & 10.0 & 17.93 &  0.58 &  0.07 &  2.1 &  160  &(27.6)&(0.79)&    12.0 &      (19.9) &  Q  &  N \\
R& 10 20 28.96 & 44 29 26.0 & 10.7 & 18.44 &  0.32 & -0.06 &  0.4 &  160  &(27.6)&(0.79)&    12.0 &      (19.9) &  Q  &  N \\
R& 10 20 28.96 & 44 29 26.0 & 11.5 & 17.80 &  0.62 & -0.09 &  1.7 &  160  &(27.6)&(0.79)&    12.0 &      (19.9) &  Q  &  N \\
 & 10 23 30.75 & 35 40 27.0 & 14.9 & 20.85 &  1.36 & -0.32 &-11.7 &  230  &(23.7)&(1.02)&         &       16.0  &  M  &  F \\
 & 10 24 13.69 & 34 57 33.1 & 11.4 & 19.52 &  1.33 &  0.63 & -5.6 &  120  &(25.4)&(0.70)&         &      (67.4) &  M  &  n \\
 & 10 32 02.14 & 39 26 14.1 &  5.1 & 18.47 &  1.01 & -0.15 & -1.5 &  190  &  30  & 0.81 &     1.2 &       42.0  &  Q  &  y \\ 
 & 10 32 41.85 & 34 48 48.2 &  7.1 & 17.37 &  0.56 &  0.06 & -3.4 &  160  &(19.0)&(0.96)&         &     ($<22$) &  G  &  C \\
 & 10 34 16.66 & 44 43 46.9 & 11.0 & 20.25 &  1.03 & -0.26 &-12.7 &  280  &  43  & 0.82 &    17.7 &        9.4  &  M  &  y \\
 & 10 35 56.22 & 37 45 02.1 & 11.5 & 17.86 &  0.42 &  0.11 &  0.7 &  210  &(55.4)&(0.60)&    17.5 &       70.0  &  Q  &  F \\
 & 10 38 08.37 & 40 21 27.2 & 10.1 & 21.37 &  1.65 &  1.44 &-11.8 &  170  &  22 &  0.89 &     0.5 &       10.0  &  M  &  Y \\
 & 10 43 25.80 & 46 08 50.3 & 11.6 & 20.39 &  0.84 & -1.32 & -7.3 &  250  &(28.3)&(0.98)&         &     (51.8)  &  ?  &  C$^{j}$ \\
 & 10 54 24.25 & 45 36 19.8 & 10.4 & 20.48 &  1.20 &  0.13 & -2.6 &  110  &(13.9)&(0.93)&         &     ($<25$) &EC1  &  C\\
 & 11 03 07.88 & 47 49 23.2 & 14.9 & 18.25 &  1.32 & -0.10 &-15.5 &  730  &(93.8)&(0.92)&         &     (14.8)  &  ?  &  C$^{j}$ \\
\hline\hline
\end{tabular}
{\caption[1]{\label{tab:tab2} Table summarising the radio and optical properties of the 
subsidiary sample of candidate 7CQ quasars. Format is the same as Table 1 with additional
references for Column 15: S, Reid et al. (1995) and Owen \& Puschell (1984).
}}
\end{center}
\end{table*}
\normalsize

\clearpage

\section{Radio observations}
\label{sec:VLA}

The observations were made on 1990 April 14, 17 and 19, and 1991 August 2 with the 
VLA in A-array, using two IFs of bandwidth 50 MHz at a mean frequency of 1.49 GHz.  
Two 4-min snapshots, separated by roughly 2 hr, were made of each source.  The flux 
density calibrator was 3C 286 for which a flux density of 14.5 Jy was assumed.

The data were calibrated in the standard way at the VLA.  Subsequent reduction was 
carried out using the AIPS package.  Where possible, 
after initial mapping and cleaning, the data were self-calibrated once for phase.  The 
Gaussian restoring beam on the full-resolution maps was circular with a FWHM of 1.4 
arcsec; lower resolution maps, for which the FWHM of the restoring beam was 4 
arcsec, were made in every case to look for larger-scale structure. 
The flux densities, positions and sizes of 
compact features were measured from the maps by fitting a Gaussian to them using the AIPS 
task IMFIT; the flux densities of more extended features were estimated using 
TVSTAT.

For A-array at 1.49 
GHz the VLA is intrinsically insensitive to structure with angular scale above 38 arcsec, a value which
may be compromised somewhat by the incomplete UV-coverage of our snapshot observations; bandwidth 
smearing also limits the accuracy of the maps at distances of $>30 ~ \rm arcsec$ from the 
pointing centre. Some of the sources observed here have structures on these 
scales, and we have utilised the 1.4-GHz NVSS survey (Condon et al. 1998) to recover reliable 
flux densities in these cases.

\section{Results}
\label{sec:results}

The results of our observations are summarised in Table 1 
(for the main sample) and Table 2 (for the subsidiary sample); radio maps of the 
extended sources in these two samples are shown in Fig.\ 3.
The optical position of each quasar candidate was examined in relation to the 
high resolution radio map; as a result a substantial number were rejected
as candidates because their APM positions (which should be accurate 
to better than 1 arcsec) were significantly offset from any plausible site for a
jet-producing nucleus. The compact features on the VLA maps have a positional
accuracy of about $\pm 0.3$ arcsec.
Further spectroscopic observations of the remainder (Paper II) were 
carried out to determine whether or not they were quasars. Details of the
current status of each object are given in Tables 1 \& 2.

\begin{figure*}
\begin{center}
\setlength{\unitlength}{1mm}
\begin{picture}(150,190)
\put(-10,-20){\includegraphics{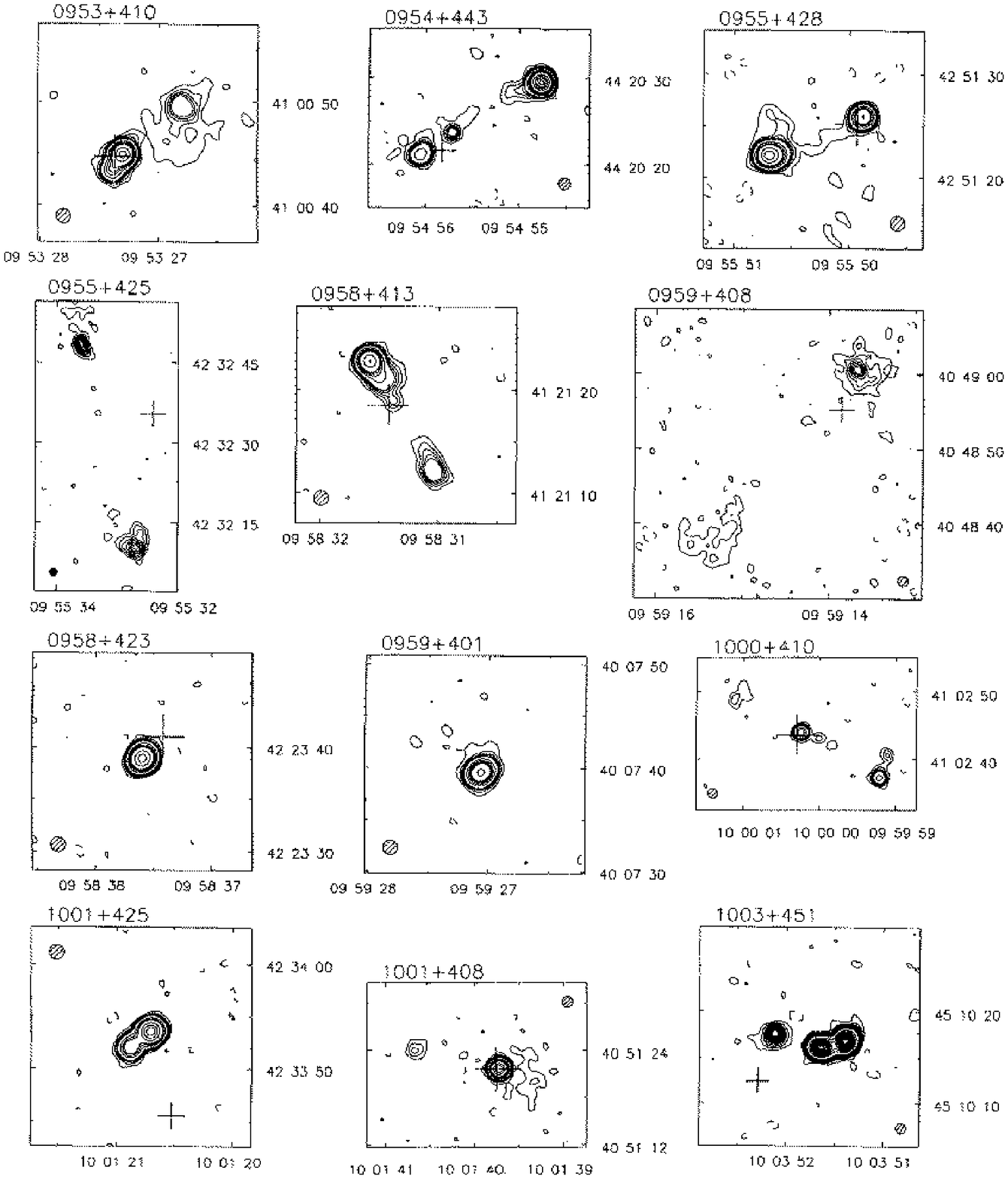}}
\end{picture}
\end{center}
{\caption[junk]{\label{fig:2a} 1.49 GHz VLA maps of the extended 
sources. Peak brightnesses and contour levels are 
given in units of mJy beam$^{-1}$; negative contours are shown dashed. 
A cross marks the APM position of the quasar 
candidate. The hatched ellipse shows the FWHM of the Gaussian restoring beam.
0953+410: peak  34.1; contours $0.4 \times (-1, 1, 2,\dots 5, 10,\dots 25, 50, 75)$. 
0954+443: peak  139; contours $1 \times (-1,1,2,\dots 5,10,\dots 25,50,\dots)$. 
0955+425: peak  5.0; contours $0.5 \times (-1, 1, 2,\dots)$. 
0955+428: peak 36.8; contours $0.5 \times (-1, 1, 2,\dots 5, 10,\dots 20, 40,\dots)$. 
0958+413: peak 30.1; contours $0.4 \times (-1, 1, 2,\dots 5, 10,\dots 25, 50, 75)$.
0958+423: peak 35.1; contours $0.4 \times (-1, 1, 2,\dots 5, 10,\dots 25, 50, 75)$. 
0959+408: peak  2.4; contours $0.3 \times (-1, 1, 2,\dots)$ . 
0959+401: peak 23.0; contours $0.4 \times (-1, 1, 2,\dots 5, 10,\dots 25, 50)$. 
1000+410: peak  7.9; contours $0.4 \times (-1, 1, 2,\dots 5, 10, 15)$.
1001+425: peak 48.0; contours $0.5 \times (-1, 1, 2,\dots 5, 10, 15, 20, 40, 60, 80)$. 
1001+408: peak 37.5; contours $0.4 \times (-1, 1, 2,\dots 5, 10,\dots 25, 50, 75)$.
1003+451: peak 83.3; contours $0.8 \times (-1, 1, 2,\dots 10, 20,\dots)$.
}}
\end{figure*}

\addtocounter{figure}{-1}

\clearpage

\begin{figure*}

\begin{center}
\setlength{\unitlength}{1mm}
\begin{picture}(150,200)
\put(-10,-20){\includegraphics{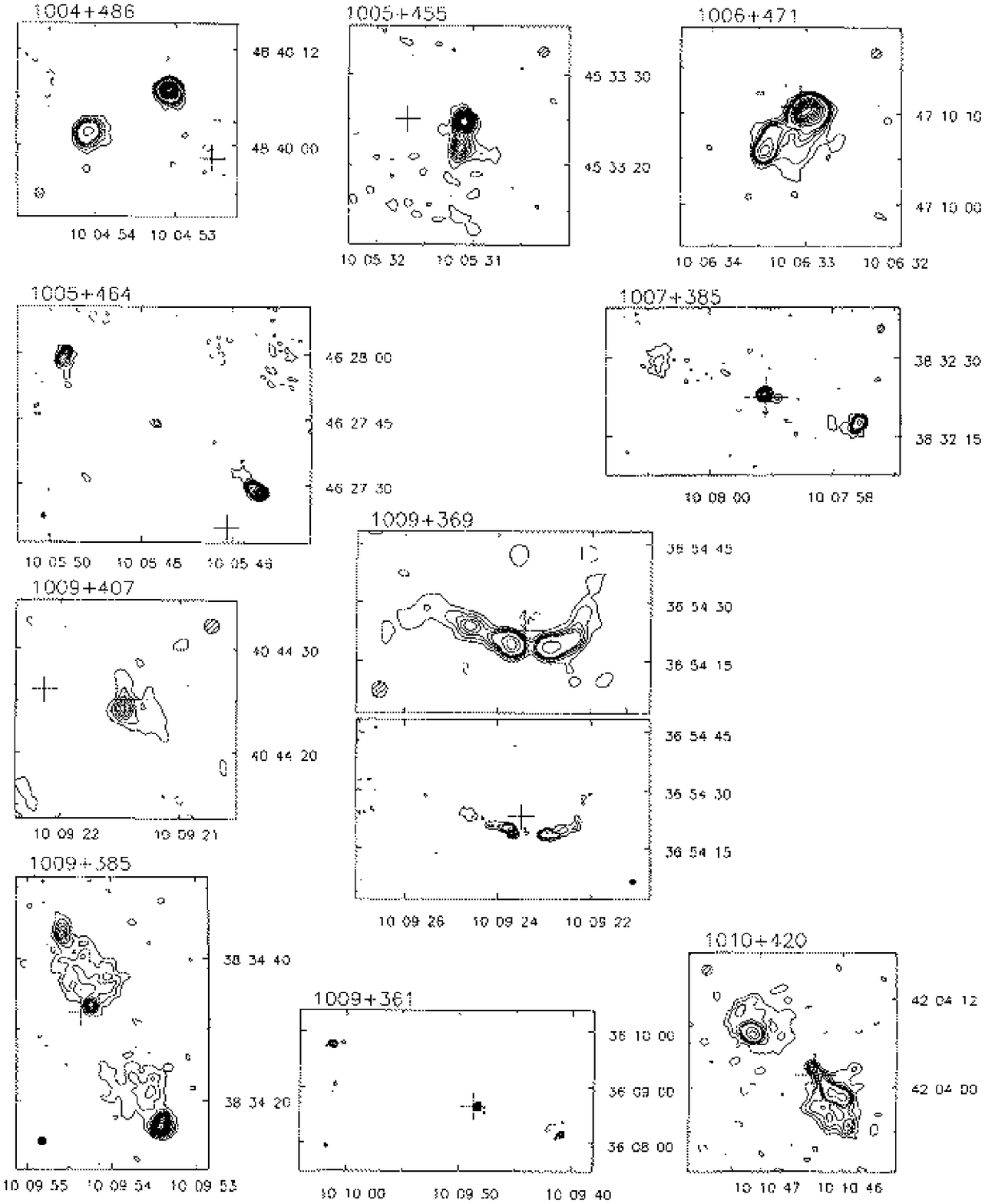}}
\end{picture}
\end{center}
{\caption[junk]{\label{fig:2b} {\bf (cont)}. 
1004+486: peak 22.0; contours, $0.4 \times (-1, 1, 2,\dots 5, 10,\dots)$. 
1005+455: peak 5.9; contours $0.5 \times (-1, 1, 2,\dots)$.
1005+464: peak 52.9; contours $0.7 \times (-1, 1, 2,\dots 10, 20,\dots)$. 
1006+471: peak 16.8; contours $0.4 \times (-1, 1, 2,\dots 5, 10,\dots)$. 
1007+385: peak 13.9; contours $0.4 \times (-1, 1, 2,\dots 5, 10,\dots)$. 
1009+407: peak  4.0; contours $0.6 \times (-1, 1, 2,\dots)$.
1009+369 (4-arcsec beam): peak 9.2; contours $0.5 \times (-1, 1, 2,\dots 5, 10, 15)$. 
1009+369 (1.4-arcsec beam): peak brightness 5.2; contours $0.4 \times (-1, 1, 2, 3, 6,\dots)$. 
1009+361 (4-arcsec beam): peak 25.0; contours $1.4 \times (-1, 1, 2,\dots 5, 10,\dots)$.
1009+385: peak 3.3; contours $0.3 \times (-1, 1, 2,\dots)$. 
1010+420: peak 7.1; contours $0.4 \times (-1, 1, 2,\dots 5, 10,\dots)$.
}}
\end{figure*}

\addtocounter{figure}{-1}

\clearpage

\begin{figure*}
\begin{center}
\setlength{\unitlength}{1mm}
\begin{picture}(150,200)
\put(-10,-20){\includegraphics{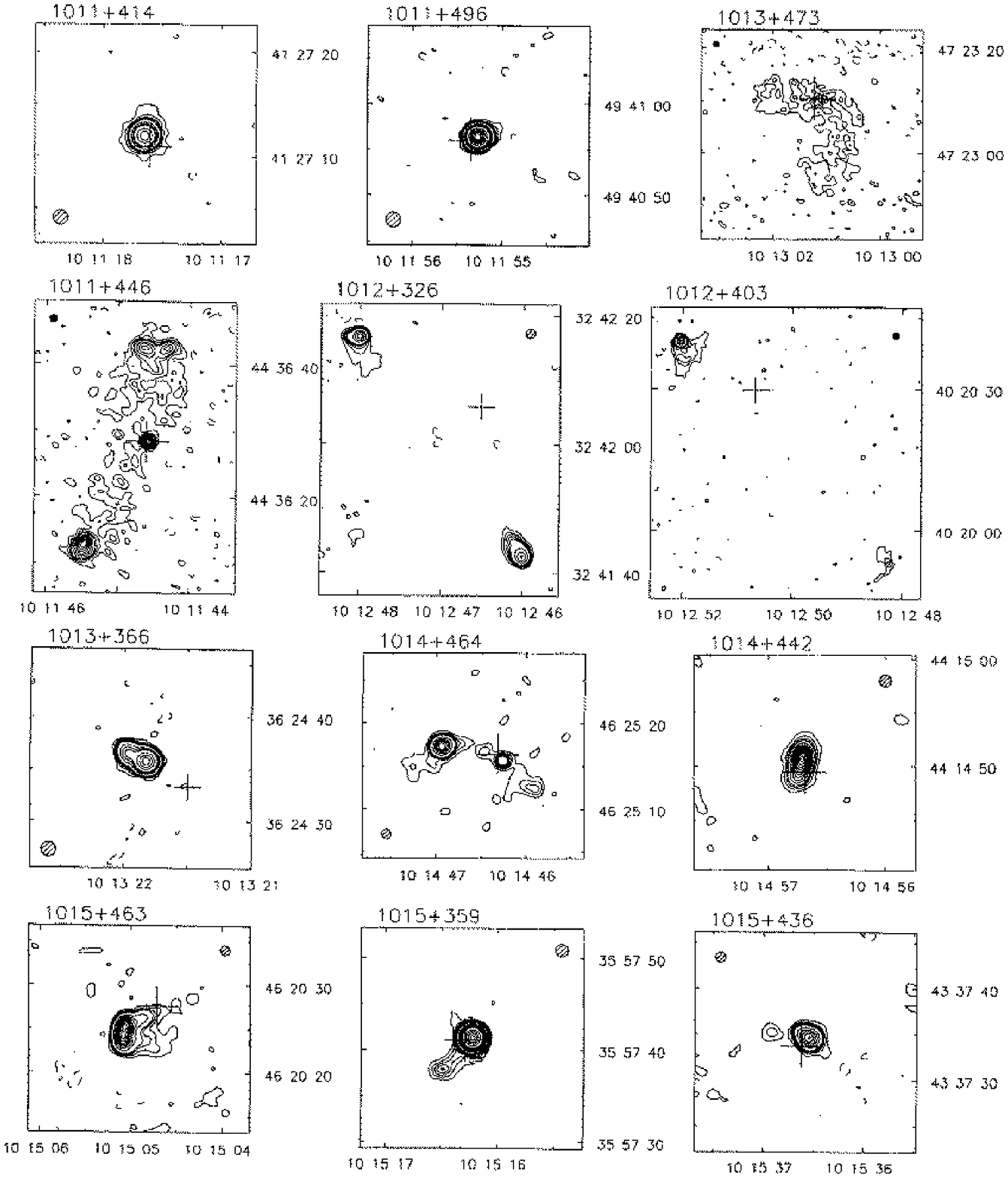}}
\end{picture}
\end{center}

{\caption[junk]{\label{fig:2c} {\bf (cont)}
1011+414: peak 36.7; contours $0.5 \times (-1, 1, 2,\dots 5, 10,\dots 20, 40, 60)$. 
1011+446: peak 4.4; contours $0.25 \times (-2, -1, 1, 2, 3, 4, 6,\dots)$. 
1011+496: peak 332; contours $2 \times (-1, 1, 2,\dots 5, 10,\dots 25, 50,\dots)$. 
1012+326: peak 19.3; contours $0.8 \times (-1, 1, 2,\dots 5, 10,\dots)$. 
1012+403: peak  8.5; contours $0.4 \times (-1, 1, 2,\dots 5, 10, 15, 20)$. 
1013+473: peak  1.3; contours $0.25 \times (-1, 1, 2,\dots)$. 
1013+366: peak 45.8; contours $0.5 \times (-1, 1, 2,\dots 5, 10,\dots 20, 40,\dots)$. 
1014+464: peak 17.4; contours $0.5 \times (-1, 1, 2,\dots 5, 10,\dots)$. 
1014+442: peak 11.3; contours $0.5 \times (-1, 1, 2,4,\dots)$. 
1015+463: peak 18.8; contours $0.5 \times (-1, 1, 2, 4, 8,\dots)$. 
1015+359: peak 556; contours $1 \times (-1, 1, 2,\dots 10, 20,\dots 100, 200,\dots)$. 
1015+436: peak 25.3; contours $0.8 \times (-1, 1, 2,\dots 5, 10,\dots)$.
}}
\end{figure*}

\addtocounter{figure}{-1}
\clearpage

\begin{figure*}
\begin{center}
\setlength{\unitlength}{1mm}
\begin{picture}(150,200)
\put(-10,-20){\includegraphics{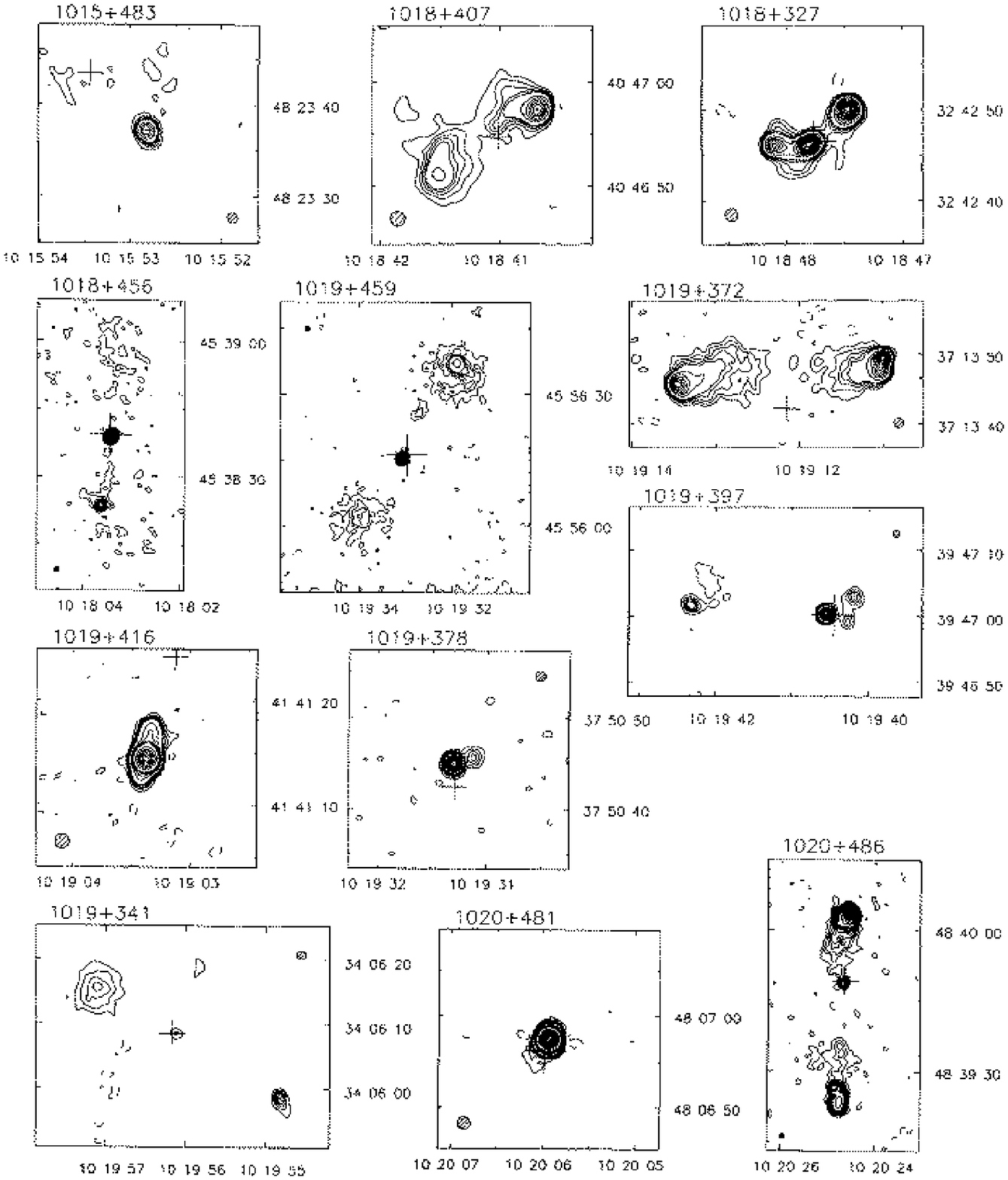}}
\end{picture}
\end{center}

{\caption[junk]{\label{fig:2d} {\bf (cont)}
1015+483: peak 13.2; contours $0.5 \times (-1, 1, 2,\dots 5, 10,\dots)$. 
1018+456: peak 46.8; contours $0.4 \times (-1, 1, 2,\dots 10, 20,\dots)$. 
1018+407: peak 28.5; contours $0.8 \times (-1, 1, 2,\dots 5, 10,\dots)$. 
1018+327: peak 53.0; contours $1 \times (-1, 1, 2,\dots 5, 10,\dots)$. 
1019+416: peak 153; contours $0.8 \times (-1, 1, 2,\dots 5, 10,\dots 25, 50,\dots)$. 
1019+372: peak 25.8; contours $0.5 \times (-1, 1, 2,\dots 5, 10\dots)$. 
1019+378: peak 10.0; contours $0.4 \times (-1, 1, 2,\dots 5, 7.5\dots)$. 
1019+459: peak 22.1; contours $0.4 \times (-1, 1, 2,\dots 5, 10,\dots)$. 
1019+397: peak 39.7; contours $1 \times (-1, 1, 2,\dots 5, 10,\dots)$. 
1019+341: peak 6.0; contours $1 \times (-1, 1, 2,\dots)$.
1020+481: peak 424; contours $1.5 \times (-2, -1, 1, 2,\dots 6, 12,\dots 30, 60,\dots)$.
1020+486: peak 31.0; contours $0.4 \times (-1, 1, 2,\dots 10, 15,\dots)$.
}}
\end{figure*}

\addtocounter{figure}{-1}
\clearpage

\begin{figure*}

\begin{center}
\setlength{\unitlength}{1mm}
\begin{picture}(150,200)
\put(-10,-10){\includegraphics{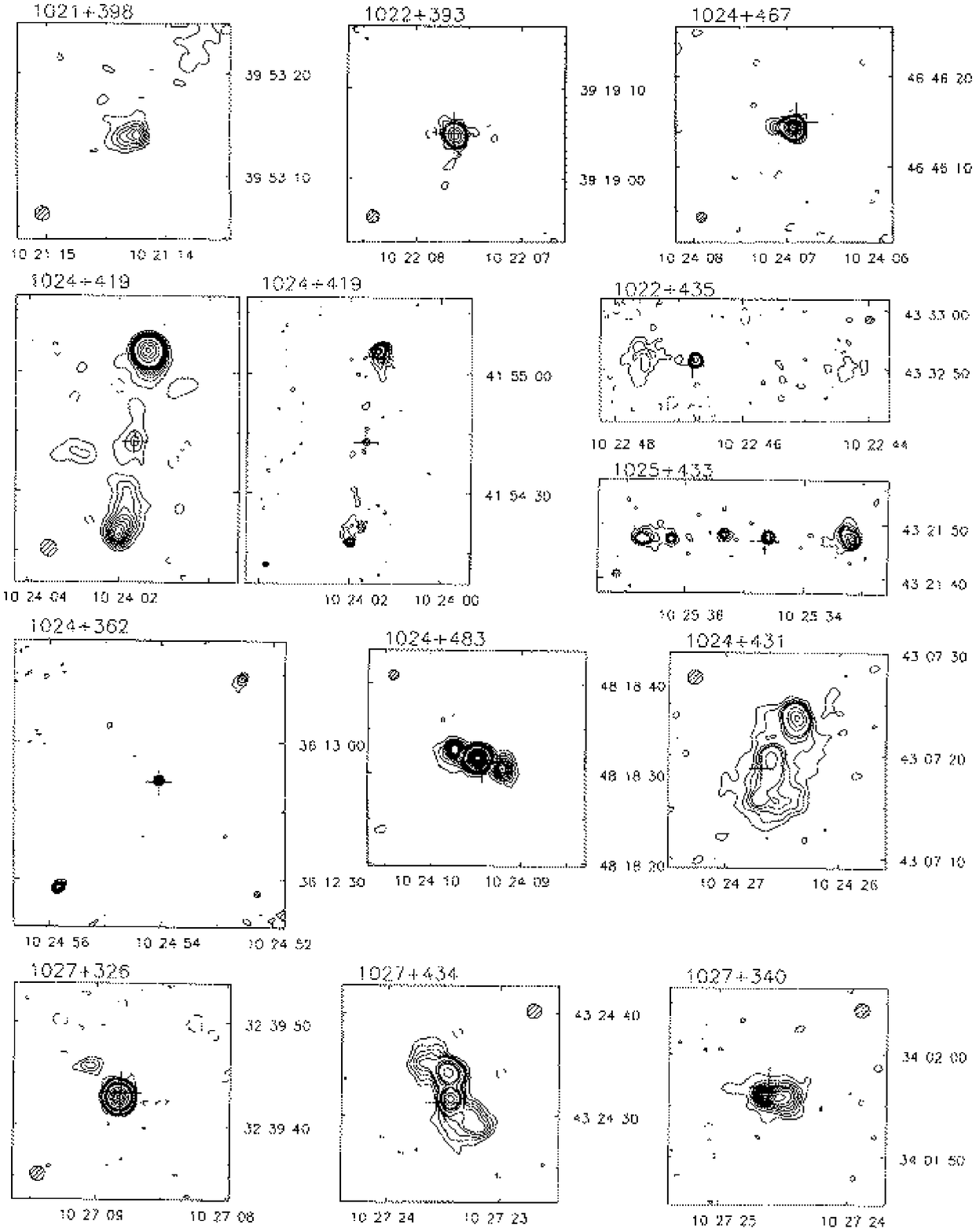}}
\end{picture}
\end{center}

{\caption[junk]{\label{fig:2e} {\bf (cont)}
1021+398: peak 6.8; contours $1 \times (-1, 1, 2,\dots)$.
1022+393: peak 17.2; contours $1 \times (-1, 1, 2, 3, 4, 8,\dots)$.
1022+435: peak 5.0; contours $0.4 \times (-1, 1, 2,\dots 5, 10)$.
1024+419 (4-arcsec beam): peak 30.5; contours $0.6 \times (-1, 1, 2,\dots 10, 20,\dots)$. 
1024+419 (1.4-arcsec beam): peak 20.0; contours $0.4 \times (-1, 1, 2,\dots 10, 20\dots)$.
1024+467: peak 14.1; contours $0.4 \times (-1, 1, 2,\dots 5, 10,\dots)$  .
1024+483: peak 156; contours $1.5 \times (-1, 1, 2,\dots 10, 20,\dots)$.
1024+431: peak 8.2; contours $0.4 \times (-1, 1, 2,\dots 5, 10,\dots)$.
1024+362: peak 5.6; contours $0.8 \times (-1, 1, 2,\dots)$.
1025+433: peak 12.6; contours $0.5 \times (-1, 1, 2,\dots 5, 10,\dots)$.
1027+326: peak 90.1; contours $0.6 \times (-1, 1, 2,\dots 5, 10\dots 25, 50,\dots)$.
1027+434: peak 34.5; contours $0.5 \times (-1, 1, 2,\dots 5, 10,\dots 20, 40, 60)$.
1027+340: peak  3.8; contours $0.3 \times (-1, 1, 2,\dots)$.
}}
\end{figure*}

\addtocounter{figure}{-1}
\clearpage

\begin{figure*}

\begin{center}
\setlength{\unitlength}{1mm}
\begin{picture}(150,200)
\put(-10,-10){\includegraphics{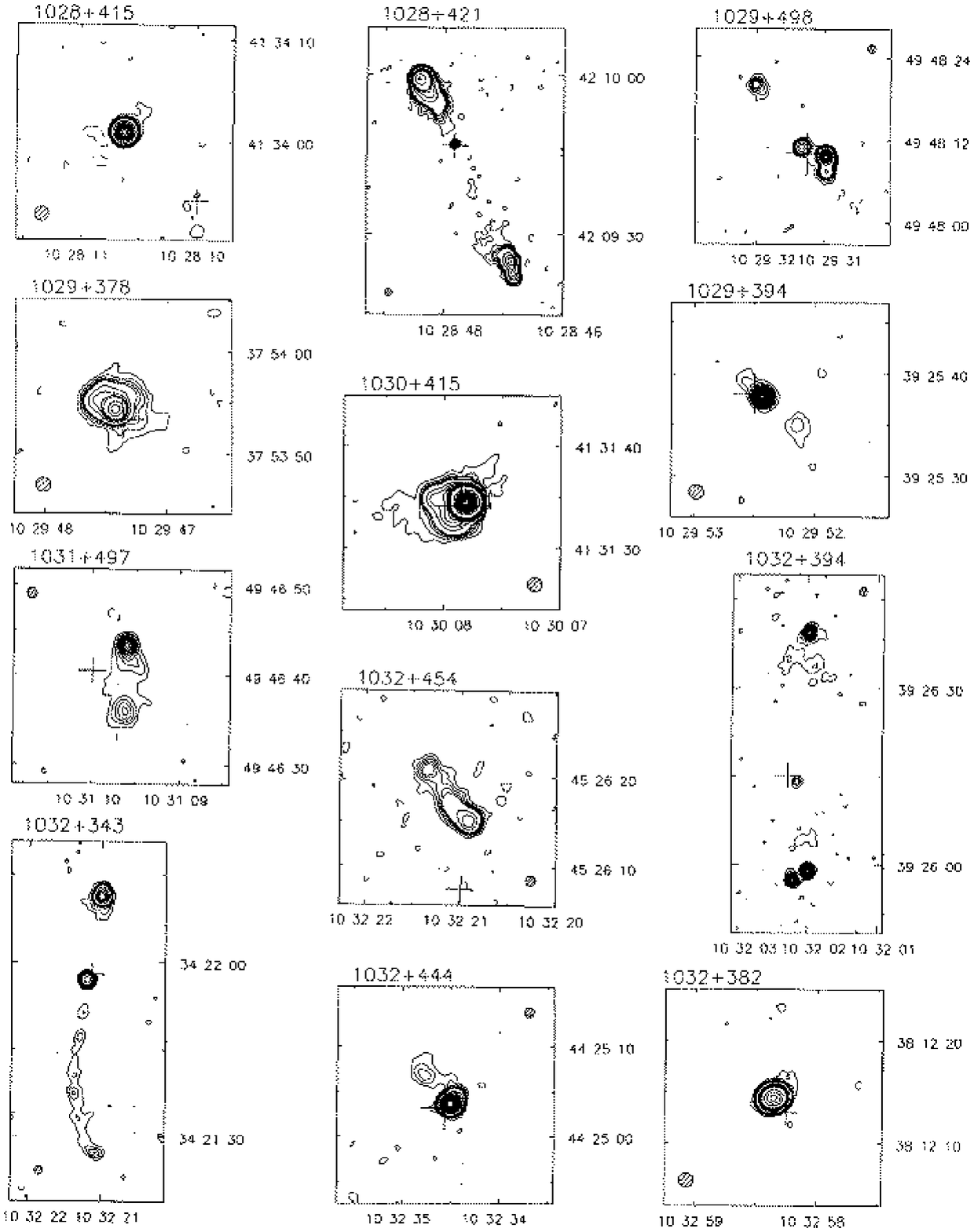}}
\end{picture}
\end{center}

{\caption[junk]{\label{fig:2f} {\bf (cont)}
1028+415: peak 23.9; contours $0.5 \times (-1, 1, 2,\dots 5, 10,\dots)$ .
1028+421: peak 30.7; contours $0.4 \times (-1, 1, 2,\dots 5, 10,\dots 25, 50, 75)$.
1029+498: peak 21.5; contours $0.5 \times (-1, 1, 2,\dots 5, 10,\dots)$.
1029+378: peak 36.1; contours $0.4 \times (-1, 1, 2,\dots 5, 10,\dots 25, 50, 75)$.
1029+394: peak 6.5; contours $0.4 \times (-1, 1, 2,\dots)$ .
1030+415: peak 486; contours $0.4 \times (-1, 1, 2,\dots 5, 10\dots 50, 150, 250,\dots)$.
1031+497: peak 6.2; contours $0.6 \times (-1, 1, 2,\dots)$.
1032+394: peak 4.4; contours $0.3 \times (-1, 1, 2,\dots)$.
1032+454: peak 5.8; contours $0.3 \times (-1, 1, 2,\dots 5, 10, 15)$.
1032+343: peak 43.1; contours $0.6 \times (-1, 1, 2,\dots 5, 10,\dots 25, 50)$.
1032+444: peak 23.4; contours $0.4 \times (-1, 1, 2,\dots 5, 10,\dots)$.
1032+382: peak 44.5; contours $0.4 \times (-1, 1, 2,\dots 5, 10,\dots 25, 50,\dots)$.
}}
\end{figure*}

\addtocounter{figure}{-1}
\clearpage

\begin{figure*}

\begin{center}
\setlength{\unitlength}{1mm}
\begin{picture}(150,200)
\put(-10,-20){\includegraphics{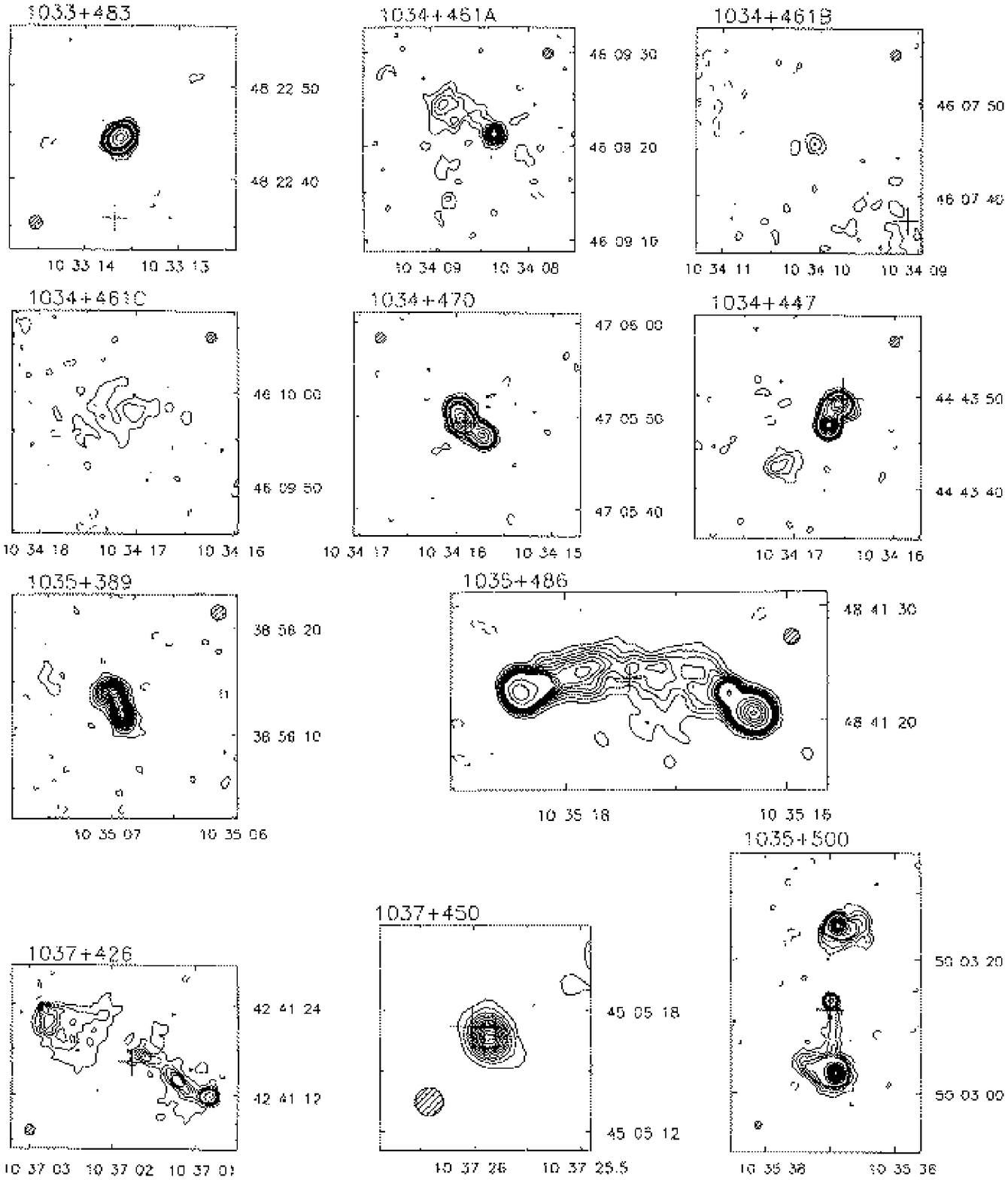}}
\end{picture}
\end{center}

{\caption[junk]{\label{fig:2g} {\bf (cont)}
1033+483: peak 51.2; contours $0.6 \times (-1, 1, 2,\dots 5, 10,\dots 25, 50,\dots)$.
1034+461A: peak 3.0; contours $0.3 \times (-1, 1, 2,\dots)$.
1034+461B: peak 1.2; contours $0.4 \times (-1, 1, 2, 3)$.
1034+461C: peak 0.8; contours $0.3 \times (-1, 1, 2)$.
1034+470: peak 51.0; contours $0.5 \times (-1, 1, 2,\dots 5, 10,\dots 20, 40,\dots)$.
1034+447: peak 16.4; contours $0.4 \times (-1, 1, 2,\dots 5, 10,\dots)$.
1035+389: peak 5.5; contours $0.4 \times (-1, 1, 2,\dots)$.
1035+486: peak 31.4; contours $0.5 \times (-1, 1, 2,\dots 10, 20,\dots)$.
1035+500: peak 23.1; contours $0.4 \times (-1, 1, 2,\dots 5, 10,\dots)$.
1037+426: peak 7.6; contours $0.5 \times (-1, 1, 2,\dots 5, 10, 15)$.
1037+450: peak 6.4; contours $0.5 \times (-1, 1, 2,\dots)$.
}}
\end{figure*}

\addtocounter{figure}{-1}
\clearpage

\begin{figure*}

\begin{center}
\setlength{\unitlength}{1mm}
\begin{picture}(150,200)
\put(-10,-20){\includegraphics{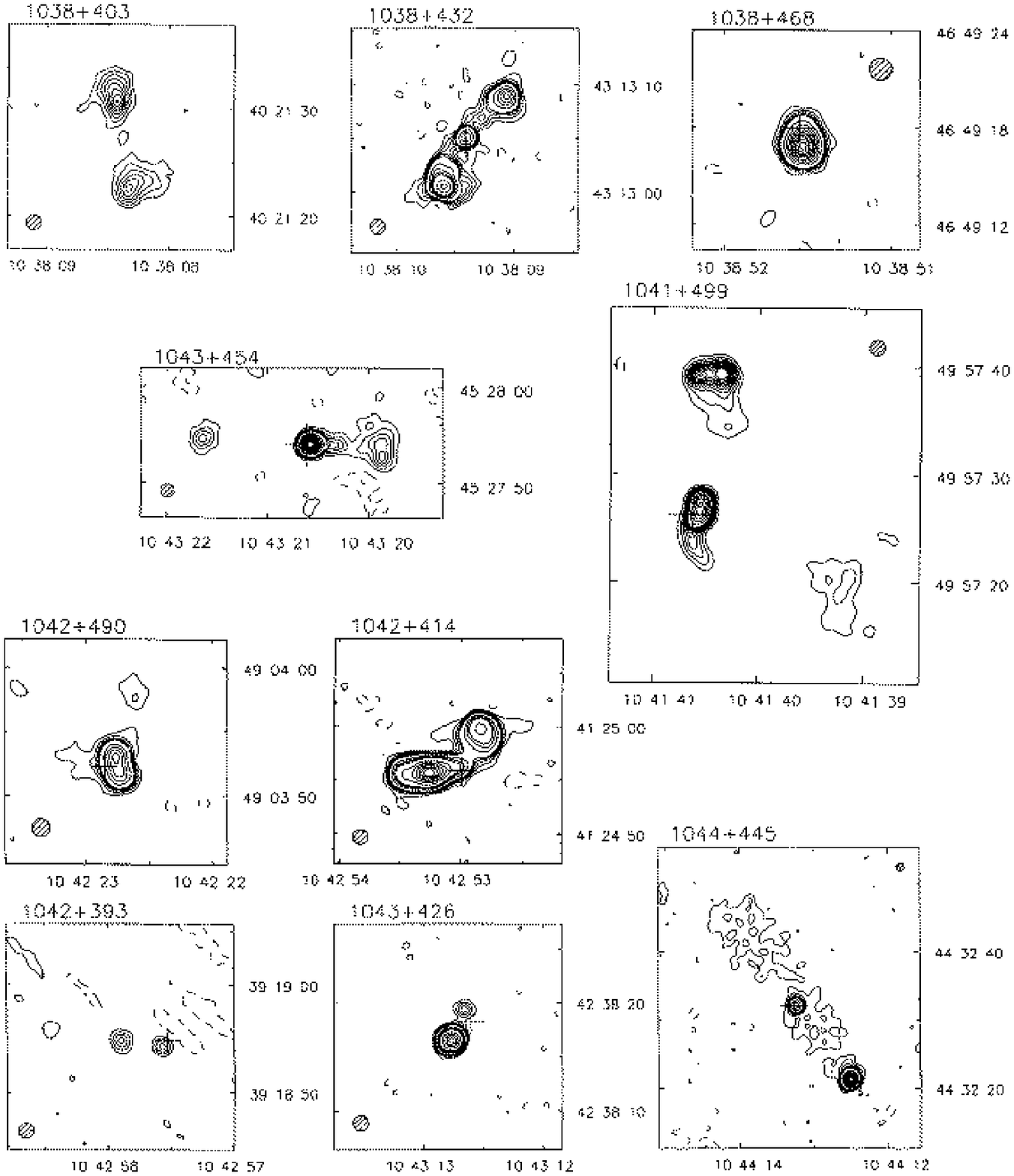}}
\end{picture}
\end{center}

{\caption[junk]{\label{fig:2h} {\bf (cont)}
1038+403: peak 3.2; contours $0.4 \times (-1, 1, 2,\dots)$.
1038+432: peak 31.2; contours $0.4 \times (-1, 1, 2,\dots 5, 10,\dots 25, 50, 75)$.
1038+468: peak 21.2; contours $0.4 \times (-1, 1, 2,\dots 5, 10,\dots)$.
1041+499: peak 21.5; contours $0.6 \times (-1, 1, 2,\dots 10, 15,\dots)$.
1042+490: peak 10.5; contours $0.4 \times (-1, 1, 2,\dots 5, 10,\dots)$.
1042+414: peak 79.2; contours $0.5 \times (-1, 1, 2,\dots 5, 10,\dots 20, 40,\dots)$.
1042+393: peak 4.4; contours $1 \times (-1, 1, 2,\dots)$.
1043+426: peak 54.8; contours $0.4 \times (-1, 1, 2,\dots 5, 10,\dots 25, 50,\dots)$.
1043+454: peak 21.8; contours $0.4 \times (-2, -1, 1, 2,\dots 5, 10,\dots)$.
044+445: peak 12.7; contours $0.3 \times (-1, 1, 2,\dots 5, 10,\dots)$.
}}
\end{figure*}

\addtocounter{figure}{-1}
\clearpage

\begin{figure*}

\begin{center}
\setlength{\unitlength}{1mm}
\begin{picture}(150,200)
\put(-10,-20){\includegraphics{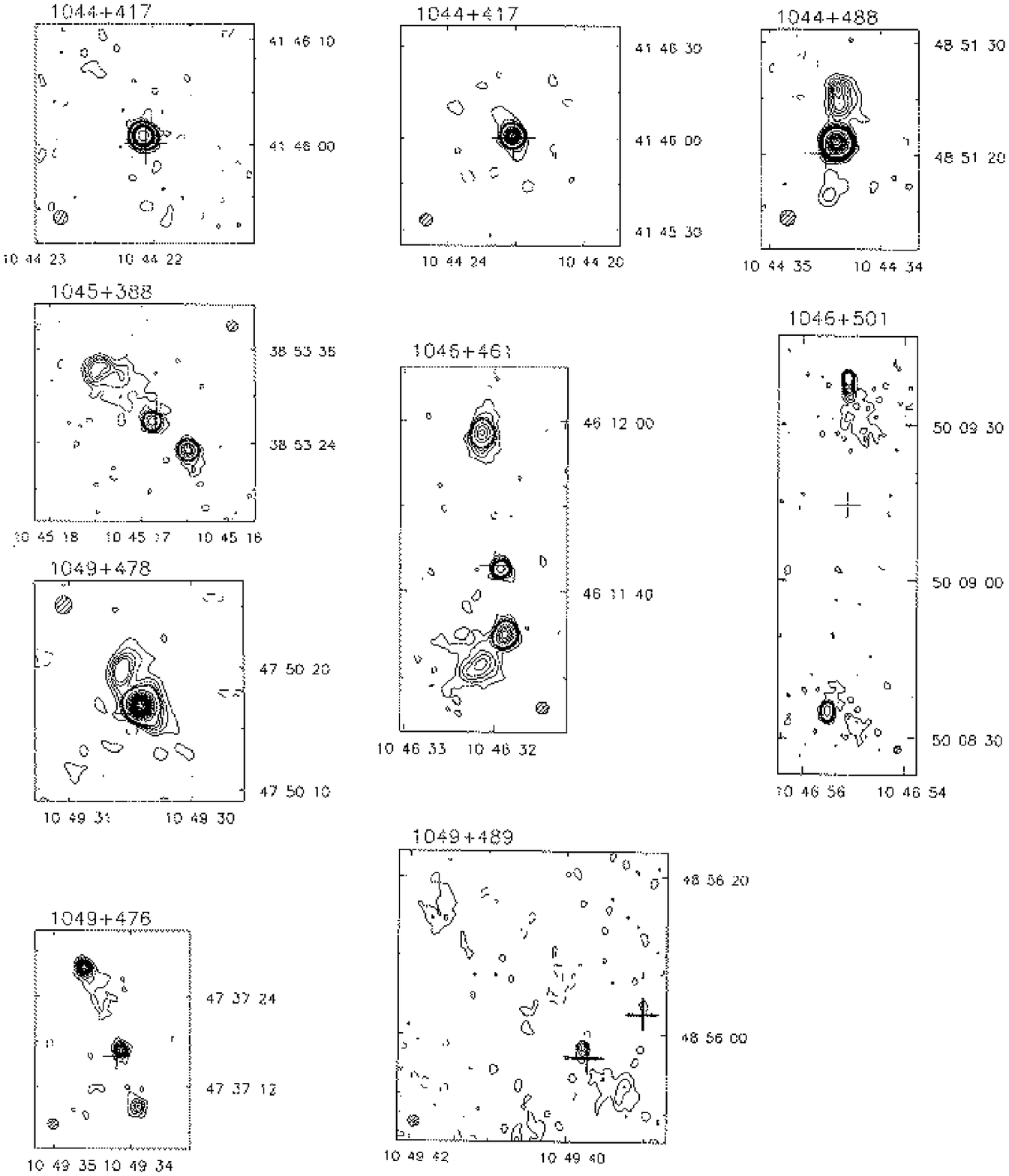}}
\end{picture}
\end{center}

{\caption[junk]{\label{fig:2i} {\bf (cont)}
1044+417 (1.4-arcsec beam): peak 26.8; contours $0.3 \times (-1, 1, 2,\dots 5, 10,\dots 30, 60,\dots)$. 
1044+417 (4-arcsec beam): peak 28.2; contours $0.7 \times (-1, 1, 2,\dots 5, 10,\dots)$. 
1044+488: peak 42.7; contours $0.5 \times (-1, 1, 2,\dots 10, 20,\dots)$.
1045+388: peak 8.9; contours $0.3 \times (-1, 1, 2,\dots 5, 10,\dots)$.
1046+461: peak 11.4; contours $0.4 \times (-1, 1, 2,\dots 5, 10,\dots)$.
1046+501: peak 4.1; contours $0.3 \times (-1, 1, 2,\dots 5, 10,\dots)$.
1049+478: peak 25.2; contours $0.4 \times (-1, 1, 2,\dots 5, 10,\dots)$.
1049+476: peak 3.9; contours $0.4 \times (-1, 1, 2,\dots)$.
1049+489: peak 2.3; contours $0.5 \times (-1, 1, 2,\dots)$.
}}
\end{figure*}

\addtocounter{figure}{-1}
\clearpage

\begin{figure*}

\begin{center}
\setlength{\unitlength}{1mm}
\begin{picture}(150,200)
\put(-10,-20){\includegraphics{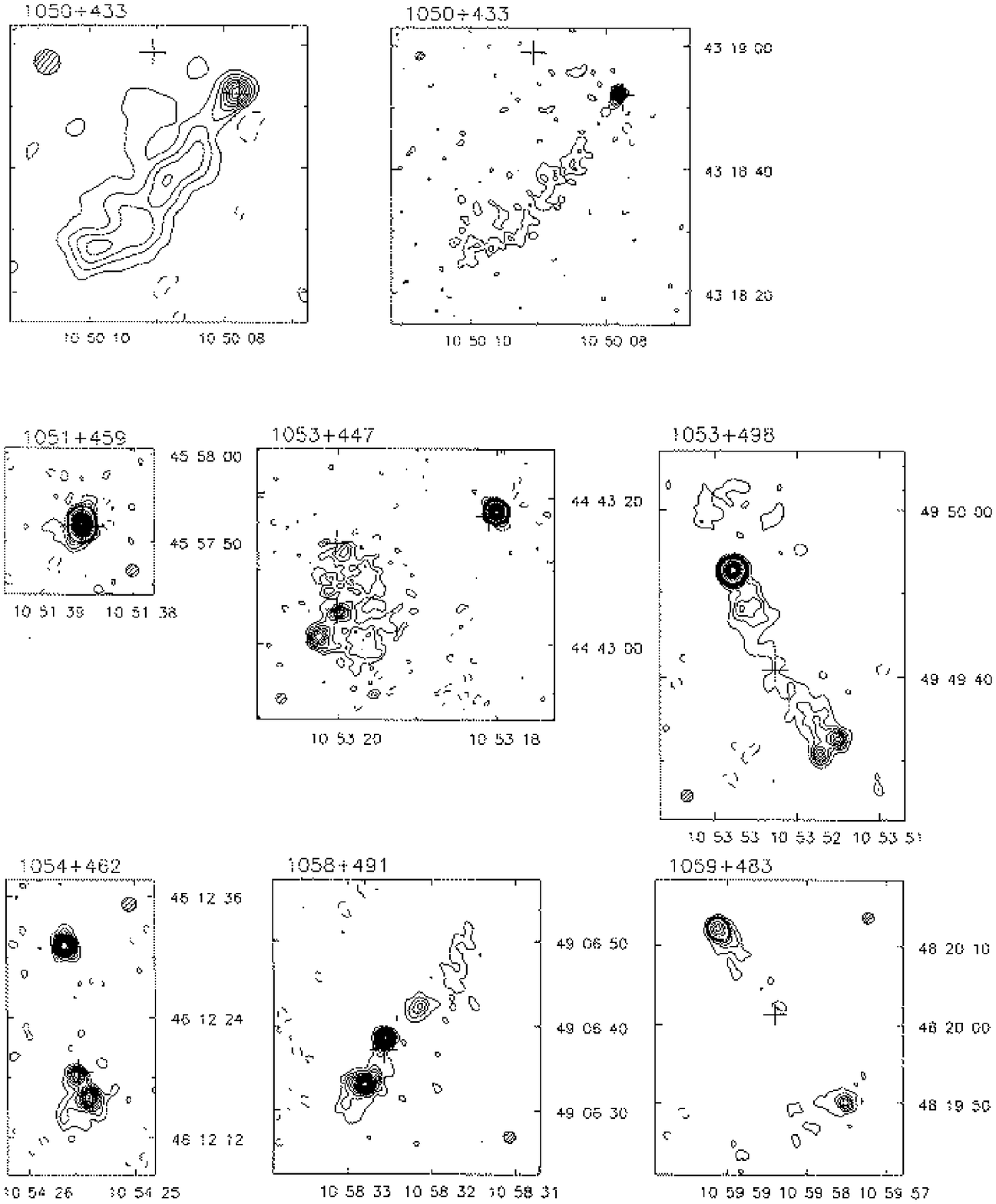}}
\end{picture}
\end{center}

{\caption[junk]{\label{fig:2j} {\bf (cont)}
1050+433 (4-arcsec beam): peak 4.5; contours $0.6 \times (-1, 1, 2,\dots)$. 
1050+433 (1.4-arcsec beam): peak 4.0; contours $0.3 \times (-1, 1, 2,\dots)$.
1051+459: peak 30.3; contours $0.4 \times (-1, 1, 2,\dots 5, 10,\dots)$.
1053+447: peak 53.0; contours $0.4 \times (-1, 1, 2,\dots 10, 20,\dots)$.
1053+498: peak 46.4; contours $0.6 \times (-1, 1, 2,\dots 10, 20,\dots)$.
1054+462: peak 3.5; contours $0.25 \times (-1, 1, 2,\dots)$.
1058+491: peak 7.6; contours $0.4 \times (-1, 1, 2,\dots)$.
1059+483: peak 10.9; contours $0.5 \times (-1, 1, 2,\dots 5, 10,\dots)$.
}}
\end{figure*}

\clearpage

\subsection{Notes on individual sources}

\begin{description}

\item[{\bf 0954+443}] (Table 2) The APM position is 2 arcsec south of the radio component which seems 
likely to be the core. The optical object has, however, been confirmed as a quasar --- the
origin of this unusually large radio/optical offset will be discussed in Paper II.

\item[{\bf 0955+425}] (Table 2) This is not included in the final version of the 7C/APM cross-matched list
because the optical object was re-classified as non-stellar on both the $E$ and $O$ POSS-I plates.

\item[{\bf 1004+486}] (Table 2)
This is not part of the final version of the 7C/APM cross-matched list
because the optical object was re-classified as non-stellar on both the $E$ and $O$ POSS-I plates.

\item[{\bf 1009+407}] (Table 2)
There are two stellar objects in the field of this source one of which,
detailed in Table 1, is 
coincident with the radio peak.
The original 7C/APM cross-match picked up the other stellar object
(which is detailed in Table 2).

\item[{\bf 1009+361}] (Table 1)
There are three radio sources in this field (with a maximum separation of 264 arcsec)
of which the central one is coincident with a quasar. The two other sources are significantly affected by bandwidth smearing 
on our VLA map but are, at most, only slightly resolved.  There is no 
evidence for any structure linking the three sources on the 151-MHz map of the area,
or on either our VLA map or the FIRST map. Inspection of the NVSS map 
does show some evidence for low-level bridge emission connecting the core
to the component to the south-west, but although the three components do 
lie roughly on a straight line, there is as yet no unambiguous
evidence that all three are related.

\item[{\bf 1013+473}] (Table 2)
This is not included in the final version of the 7C/APM cross-matched list
because the optical object was re-classified as non-stellar on both the $E$ and $O$ POSS-I plates;
this object has been confirmed as a galaxy by optical spectroscopy (paper II).

\item[{\bf 1019+416}] (Table 2)
This is not included in the final version of the 7C/APM cross-matched list
because the optical object was re-classified as non-stellar on the $E$ POSS-I plate, and
was confused on the $O$ plate.

\item[{\bf1019+378 }] (Table 1)
The position of the stellar object is 2.5 arcsec south of the radio component 
which seems likely to be the core; further astrometric and spectroscopic work is 
required to confirm whether this object is really associated with the source.

\item[{\bf 1022+435}] (Table 1)
There is some diffuse structure to the east of the core which has not been 
adequately mapped by our VLA observations.  We have therefore preferred 
to tabulate the flux density 
at 1.4 GHz from the NVSS survey.

\item[{\bf 1024+419}] (Table 1)
The low-resolution map shows possible low-brightness wings extending away from 
the axis of the source. The tabulated 1.4-GHz NVSS flux density 
significantly exceeds the value ($S_{1490}=76 ~ \rm mJy$) estimated from our map.
 
\item[{\bf 1024+349}] (Table 2)
The source was classified as extended (40 arcsec) in 7C; no definite extended structure was detected by 
our VLA observations but comparison with the NVSS data shows that this is
simply because the surface brightness is too low. Extended structure is clearly apparent in the
FIRST map.
The original stellar object can be rejected on positional grounds; however 
there is a very faint ($E = 19.7$) optically-unresolved object coincident with the peak of the 
source on our VLA map (RA 10 24 13.69, DEC 34 57 33.1), but this object is too red to appear on the $O$ plate and may be
a galaxy.

\item[{\bf 1028+415}] (Table 1)
There is marginal evidence for a faint extension $\sim2 ~ \rm arcsec$ in p.a. 315$^{\circ}$ although it
remains possible that this is an artefact of imperfect calibration.
 
\item[{\bf 1029+498}] (Table 1)
The most easterly of the three bright regions in the southern 
component seems most likely to be the core on positional grounds since the 
other radio sub-components are more than 3 arcsec from the optical position.

\item[{\bf 1032+394}] (Table 2)
The position of the optical object (which the APM analysis finds to be extended on both the $E$ and $O$ 
POSS-I plates) is 1.7 arcsec east and 1.4 arcsec north of that of the core; further astrometric and 
spectroscopic work is required to confirm whether this object is really associated with 
the source.

%\item[{\bf 1032+348}] 
%This is not included in the final version of the 7C/APM cross-matched list
%because the optical object was re-classified as non-stellar on both the $E$ and $O$ POSS-I plates;
%this object has been confirmed as a galaxy by optical spectroscopy (paper II).

\item[{\bf 1034+461}] (Table 1)
There are three radio sources in this field; there is no evidence that any of these
are associated with the stellar object.

\item[{\bf 1034+447}] (Table 2)
The original candidate ID can be rejected on positional grounds (it is not shown on the radio map). 
The cross shows the position of a very faint ($O = 21.8$) object 
detected only on the $O$ POSS-I plate (RA 10 34 16.55, DEC 44 43 49.4); 
it lies close to the northernmost peak of the radio 
source and is a plausible, but not compelling, optical ID.

\item[{\bf 1042+393}] (Table 1)
There is a possible very weak component 10 arcsec east of the core; 
however the map is of poor quality and, especially as there is no
corresponding feature on the FIRST map, it may not be real
The flux density from the NVSS survey is preferred in {\mbox Table 1}.

\item[{\bf 1049+489}] (Table 1)
The noise on this map is high because of  residual sidelobe emission from a bright source about 
4 arcmin to the south-west of the 7C radio source, and it is thus not clear whether the
component in the north-east of the map is real. There are two stellar objects on the POSS-I $E$ plate 
near to the 7C source. The brightest of these is a redshift $z = 0.478$ quasar
known in the literature as 5C2.10, but a comparison of our APM finding chart and the optical image 
of the field presented by Ellingson \& Yee (1994) shows that the position they give
for this quasar (RA 10 49 41.00, Dec 48 55 53.0) is in error, and that our
new APM position (10 49 39.06, 48 56 01.9) should be preferred. Moreover, our radio map shows
that this quasar is not likely to be the true ID of the 7C (and 5C)
radio source; this quasar, which lies in a rich cluster, should be re-designated as a 
radio-quiet object. There is 
a second stellar object which is significantly closer to the radio core position
($\Delta \rm RA = 0\!\stackrel{s}{.}\!03$, $\Delta \rm Dec = 1.4 ~ arcsec$). 
It has an $E$ magnitude of 19.1 but is not visible on the $O$ plate
so that $O-E > 2.4$. It corresponds to Object 391 of Ellingson \& Yee
who failed to secure a redshift for it so it it is not yet clear 
whether the radio source is identified with a spatially-compact
galaxy, which may be in the same cluster as the known quasar, or whether it is a background or foreground 
quasar with an unusually red colour.   
The radio source was also detected in B-array VLA observations at 1.4 GHz 
by Hutchings et al. (1996).

\item[{\bf 1050+460}] (Table 1)
This candidate was omitted from the VLA programme because the
optical object was classified as non-stellar on the $E$ POSS-I plate; subsequent
spectroscopy (paper II) has revealed that this object is 
nevertheless a quasar.

\item[{\bf 1050+433}] (Table 1)
The original candidate ID can be rejected on positional grounds.  
There is, however, a galaxy with an $E$ magnitude of 14.7 coincident with the core of 
this radio trail source ($\Delta \rm RA = -0\!\stackrel{s}{.}\!02$, 
$\Delta \rm Dec = 0.0 ~ arcsec$).

\item[{\bf 1053+447}] (Table 1)
It is not clear whether the compact source to the north-west whose 
radio position is given in Table 1 is related to the extended source in the south-east.  There 
are three optical objects in the field which may be related; their positions, $O$ 
magnitudes, and colours ($O-E$) are as follows:

\begin{tabular}{lllll}
&&&&\\
A & 10 53 18.08 & 44 43 17.3 & 19.51 & 0.30 \\
B & 10 53 20.00 & 44 43 13.6 & 19.66 & 0.89 \\
C & 10 53 20.03 & 44 43 04.4 & 19.70 & 0.71 \\
\\
\end{tabular} 

Object A, which is stellar on the $E$ plate and confused on the $O$ plate, is very close to 
the peak of the compact source and is a good ID for this component, and possibly
all the radio structure.  Object B is stellar on both plates but appears to be unrelated to 
any of the radio emission.  Object C, which is 
extended on the $E$ plate and stellar on the $O$, lies close to one of the bright peaks in the 
extended source, and is a plausible separate ID for the extended source in the south-east.

%\item[{\bf 1054+462}] (Table 1)
%This object was confused on the POSS-I $O$ plate.

\end{description}

\section{The 7CQ sample and selection effects}
\label{sec:discussion}

Full discussion of the 7CQ sample is left to future papers.
We concentrate here on analysing the possible effects
of the selection criteria outlined in Section 2. We exclude from this
discussion the 21 radio sources with $S_{151} \leq \rm 0.1 ~ Jy$ 
since these lie below our adopted radio flux density limit.
We also exclude objects which are associated with
other types of optical object, namely the 5 confirmed galaxies (category `G')
and the one BL Lac object (`BL'), and radio sources for which any proposed APM 
ID can be rejected on positional grounds (the 40 objects
in category `M') and/or spectroscopic grounds (the 9 proposed IDs which turned out to be stars, `S'). 

This leaves the following objects in Tables 1 and 2:
70 confirmed quasars (`Q'); 27 quasar candidates, i.e., objects verified as good 
optically-unresolved IDs which currently lack spectroscopic confirmation (`QC');
13 objects which are optically unresolved but which are not yet verified as good
IDs, and which in addition lack optical spectra (`?'); and 36 objects 
which are optically unresolved on only one of the two POSS-I plates 
(categories `EC1', `EC2', `OC0', `OC1' and `OC2').
The sky positions of these objects are shown in Fig.\ 4.
Taking category `Q' and `QC' objects together as our 7CQ sample we will argue that
any further missing quasars constitute only a small and unbiassed fraction of those meeting well-defined 
radio and optical flux density criteria over the sky region delineated in Fig.\ 4. 

%  There will eventually be a Section 5.1 in which we deal with the S_151 limits
%  here summarise what the final criteria are once we know the deal with S_151
%

\begin{figure*}
\begin{center}
\setlength{\unitlength}{1mm}
\begin{picture}(150,150)
\put(170,0){\includegraphics{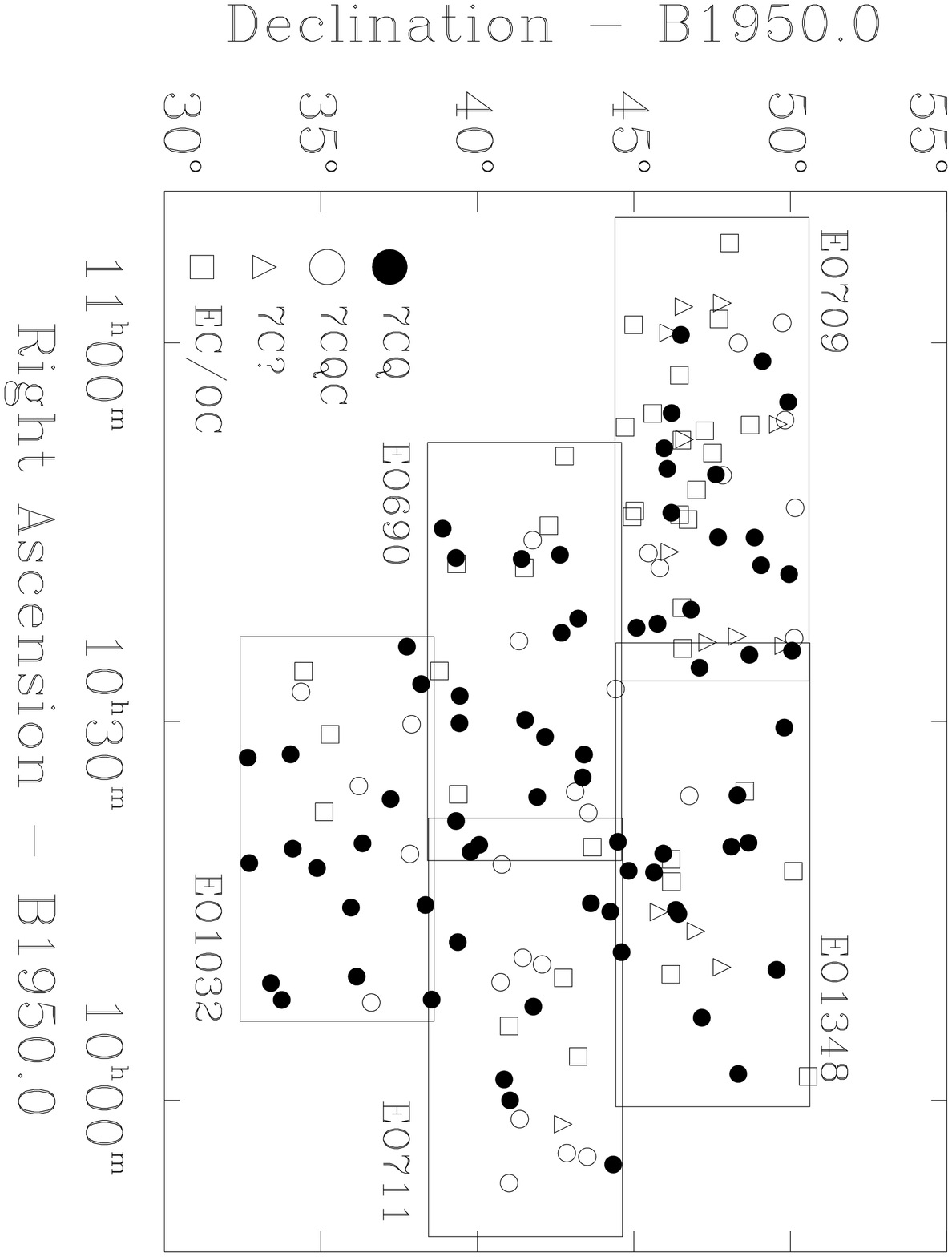}}
\end{picture}
\end{center}
{\caption[junk]{\label{fig:skypos} The sky locations of the 
following objects from Tables 1  and 2 with $S_{151} > 100 ~ \rm mJy$ (see Table 1 caption for full details of these
categories): confirmed quasars (`Q' -- filled circles); candidate quasars
(`QC' -- open circles); possible quasars (`?' -- open triangles); and
objects which have a non-stellar APM category on one plate 
(`EC1', `EC2', `OC0', `OC1', `OC2' -- open squares). 
The sky coverage of the individual POSS-I plate pairs
are illustrated by the rectangular boxes.

%
% Scale symbol size by flux?
%

}}
\end{figure*}

\subsection{Radio-loud quasars missing from Tables 1 and 2}

We consider here ways in which radio-loud quasars above the 
$S_{151}$ limit may have been excluded from Tables 1 and 2. 

First, we investigate the 
imposition of an $r$ cutoff during our sample
selection. A simple statistical analysis shows why this selection criterion was a practical necessity.
Data from the final APM analysis were used to assess the 
probability of a random coincidence between a radio source and a stellar object in a 
particular colour range. The surface density of random objects was estimated by assuming 
that the objects with $r$ greater than 20 arcsec are, indeed, chance associations.  The results are summarised in 
Table 3 which lists the observed numbers and those expected by chance with $r$ in the 
range 0--5, 5--10 and 10--15 arcsec, expressed as a function of APM colour ($O-E$) 
in the range $<$1, 1--2, 2--3 and $>$3 (see $O ~ \& ~ E$ rows in Table 3).
It can be seen that there are significant numbers of true associations amongst the 
objects with $O-E <1$ even out to 15 arcsec.  However, for the redder objects the 
proportion of chance associations increases rapidly.  It is clear from this analysis that, if 
we wish to obtain a sample of SSQs which lacks a strong colour bias, whilst at 
the same time avoiding excessive numbers of random associations, we should select 
samples of objects with an $r$ cutoff of at most 10 arcsec. This was the motivation behind
our original survey, and our first set of VLA follow-up observations. It then became clear
(as discussed in Section 5.2) that red objects could be excluded
with little risk of excluding quasars, and we therefore constructed a 
subsidiary sample of candidates with $10 < r < 15 ~ \rm arcsec$ and blue optical colours. 
Information on these objects can be found in {\mbox Table 2}.

\scriptsize
\begin{table}
\begin{center}

\begin{tabular}{l|r|r|r|r|r}
\hline\hline
\mc{1}{c|}{$O-E$} & \mc{1}{c|}{$< 1$} & \mc{1}{c|}{1-2} & \mc{1}{c|}{2-3} & 
\mc{1}{c|}{$>3$} & \mc{1}{c|}{Total} \\
\hline
\mc{6}{c}{Radio-optical separation 0--5 arcsec} \\
\hline
Observed $O$ \& $E$ & 51 & 26 & 14 &  7 &  98 \\
Expected $O$ \& $E$ &  2 & 11 & 10 &  3 &  26 \\
Observed $O$        & 10 & 40 &  0 &  0 &  50 \\
Expected $O$        &  1 & 28 &  0 &  0 &  29 \\
Observed $E$        &  0 & 16 & 31 &  4 &  51 \\
Expected $E$        &  0 & 12 & 14 &  2 &  28 \\
\hline
\mc{6}{c}{Radio-optical separation 5--10 arcsec} \\
\hline
Observed $O$ \& $E$ & 31 & 40 & 28 & 10 & 109 \\
Expected $O$ \& $E$ &  7 & 33 & 30 &  8 &  78 \\
Observed $O$        & 12 & 80 &  0 &  0 &  92 \\
Expected $O$        &  3 & 84 &  0 &  0 &  87 \\
Observed $E$        &  0 & 28 & 32 & 11 &  71 \\
Expected $E$        &  0 & 34 & 40 &  5 &  79 \\
\hline
\mc{6}{c}{Radio-optical separation 10--15 arcsec} \\
\hline
Observed $O$ \& $E$ & 24 & 60 & 47 & 11 & 142 \\
Expected $O$ \& $E$ & 11 & 55 & 49 & 13 & 128 \\
Observed $O$        & 14 &133 &  2 &  0 & 149 \\
Expected $O$        &  8 &142 &  2 &  0 & 152 \\
Observed $E$        &  0 & 55 & 55 &  8 & 118 \\
Expected $E$        &  0 & 58 & 68 &  8 & 134 \\
\hline\hline
\end{tabular} 
{\caption[junk]{\label{tab:flux} Statistics of the observed radio-optical separations 
in comparison with matches expected from random associations 
between radio sources and objects categorised as stellar by the 
APM analysis. Objects detected on both $O$ and $E$ plates are 
considered separately from those detected on just one of the plates. In this latter case we
have determined colours by assuming that the object lies just below the magnitude limit 
of the plate on which it is not detected by the APM analysis.
}}
\end{center}
\end{table}

\normalsize

Fig.\ 5 suggests that we are not missing a significant
number of quasars with $r > 10 ~ \rm arcsec$. The evidence from our subsidiary 7CQ sample 
(e.g. objects in Table 2) is that there are very few, although not zero, quasars with $r > 10 ~ \rm arcsec$, 
but that the number drops off so rapidly with $r$ that very few are expected with
$r > 15 ~ \rm arcsec$. It is extremely difficult to model this source of incompleteness
because the value of $r$ is likely to be a complicated function of
$S_{151}$ (due to increasing inaccuracy in the 7C position at fainter values of
$S_{151}$), and the internal structure of the source --- specifically on
how well the low-resolution low-frequency position matches the position of the
compact radio nucleus. Note the discussion of 1053+447 in Section 4.1 as an example
of the difficulties that can be involved. 
At the $S_{151} > 0.5 ~ \rm Jy$ level the dominant contribution is from
`real' offsets as the random errors
in 7C position are $\stackrel{<}{_\sim} 2 ~ \rm arcsec$ (McGilchrist et al. 1990).
The results of the 7C Redshift Survey (Willott et al. 1998b) provide empirical evidence that
$\ll 10$ per cent of this population have real offsets which are large
enough to cause problems: all 18 SSQs in this {\it complete} survey 
(with a flux density limit $S_{151} > 0.5 ~ \rm Jy$) have $r < 10 ~ \rm arcsec$. 
In absolute terms the largest real offsets are likely to occur for the 
quasars with the largest radio angular sizes, which will themselves 
tend to be in the high flux density samples. However, even amongst the 41 quasars 
with $S_{151} > 12 ~ \rm Jy$ in the revised 3C sample (Laing et al. 1983) there are 
none for which $r > 10 ~ \rm arcsec$. That there are likely to be very few 
quasars with very large angular sizes in the 7CQ sample is further confirmed by the work of Dingley (1990). 
He searched for the optical counterparts of radio-faint ($0.4 \leq S_{151} \leq 1.0 ~ \rm Jy$), 
large angular size ($1.5 \leq \theta \leq 3 ~ \rm arcmin$) sources across $0.33 ~ \rm
sr$ of the 7C survey (including almost all of the 7CQ area shown in Fig. 4). 
VLA and optical spectroscopic follow-up (Dingley 1990; Cotter, Rawlings \& Saunders 1996)
has shown that the areal density of quasars obeying these radio criteria (and with
optical magnitudes above the POSS-I limits) is $\stackrel{<}{_\sim} 10 ~ \rm sr^{-1}$, and few if
any are therefore expected in the 7CQ sample. 
At the lowest flux densities ($100 \leq S_{151} \leq 200 ~ \rm mJy$) the effects of the 
random positional errors will dominate over the real offsets. The standard errors in the positions at 
these flux density levels are in the range $5 - 10 ~ \rm arcsec$ (Pooley, Waldram
\& Riley 1998). The search radius of $10 ~ \rm arcsec$ can then be as much as one 
standard deviation so that up to 33 per cent of the quasars could lie outside this range.
This potential source of incompleteness must be taken into account alongside the
decreasing fraction of the 7CQ area which is complete to the faintest $S_{151}$ levels
(see {\mbox Fig. 2}). Because the positional errors are random this can be treated
as an additional decrease in the effective area of the 7CQ survey at 
$S_{151} \stackrel{<}{_\sim} ~ \rm 200 mJy$.

\begin{figure*}
\begin{center}
\setlength{\unitlength}{1mm}
\begin{picture}(150,150)
\put(200,0){\includegraphics{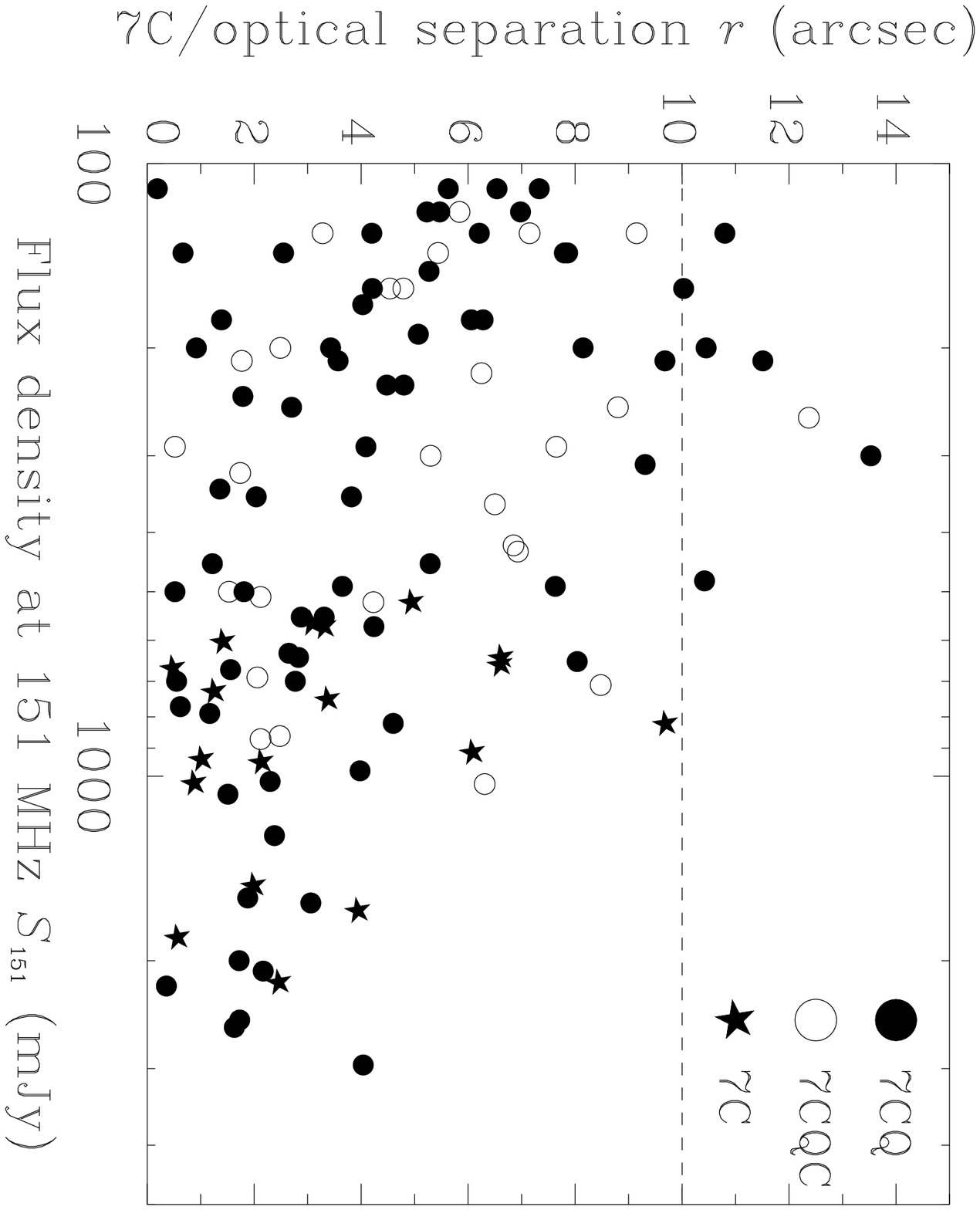}}
\end{picture}
\end{center}
{\caption[junk]{\label{fig:s151r} 
The angular separation $r$ between the 7C radio source
position and the APM optical position plotted against
the total 7C flux density. The filled circles represent confirmed quasars (category
`Q'), the open circles, candidate quasars (category `QC') and the stars, steep-spectrum
quasars from the flux-limited 7C Redshift Survey (Willott et al. 1998b).
Very few quasars lie above the dotted line which shows the $r$ cutoff used in the
original 7CQ sample.
}}
\end{figure*}

A second way in which quasars may be missing from Tables 1 and 2 results from our
insistence that quasar candidates be detected on both the $O$ and $E$ POSS-I plates ---
a question inextricably linked with the limiting optical magnitude of the 7CQ survey. 
We adopted this selection criterion  because of the large
number of 7C/APM cross-matches in which the object appeared on only
one of the plates; a summary of the numbers involved is included in
Table 3. Such objects are optically faint and lie close to the magnitude limit of the plates on which 
they were detected. The areal density of such
objects is high, and as a consequence there is a high chance that a given association 
is simply a chance projection. Nevertheless, inspection of Table 3 
shows that there is still a significant number of genuine IDs on just the
$E$ plate. However, since these objects have to be red [$(O - E) > 1.5$]
to avoid detection on the $O$ plate, most and possibly all are probably
radio galaxies, and we will not consider them further; 
the reasons for this will be discussed in Section 5.2. The 
7C/APM cross-matches amongst objects detected optically on just the $O$ plate 
is also well above the number expected by chance. These blue IDs {\em will} 
include a significant number of genuine radio-loud quasars.
We have not yet followed up these objects either with the VLA or spectroscopically. 
Although missing from Tables 1 and 2,
the existence of these quasars does not affect the completeness of the
7CQ sample provided a magnitude of $E \approx R \approx 20$ is adopted as 
the appropriate optical limit for the survey.

\subsection{Radio-loud quasars amongst objects in Tables 1 and 2 but not followed-up}

We consider here selection effects arising from incomplete VLA and optical follow-up of the
objects in Tables 1 and 2.
To illustrate this discussion we plot in Fig.\ 6 the $E$ magnitude versus
the $O-E$ colour for all objects in these tables.

\begin{figure*}
\begin{center}
\setlength{\unitlength}{1mm}
\begin{picture}(150,150)
\put(200,0){\includegraphics{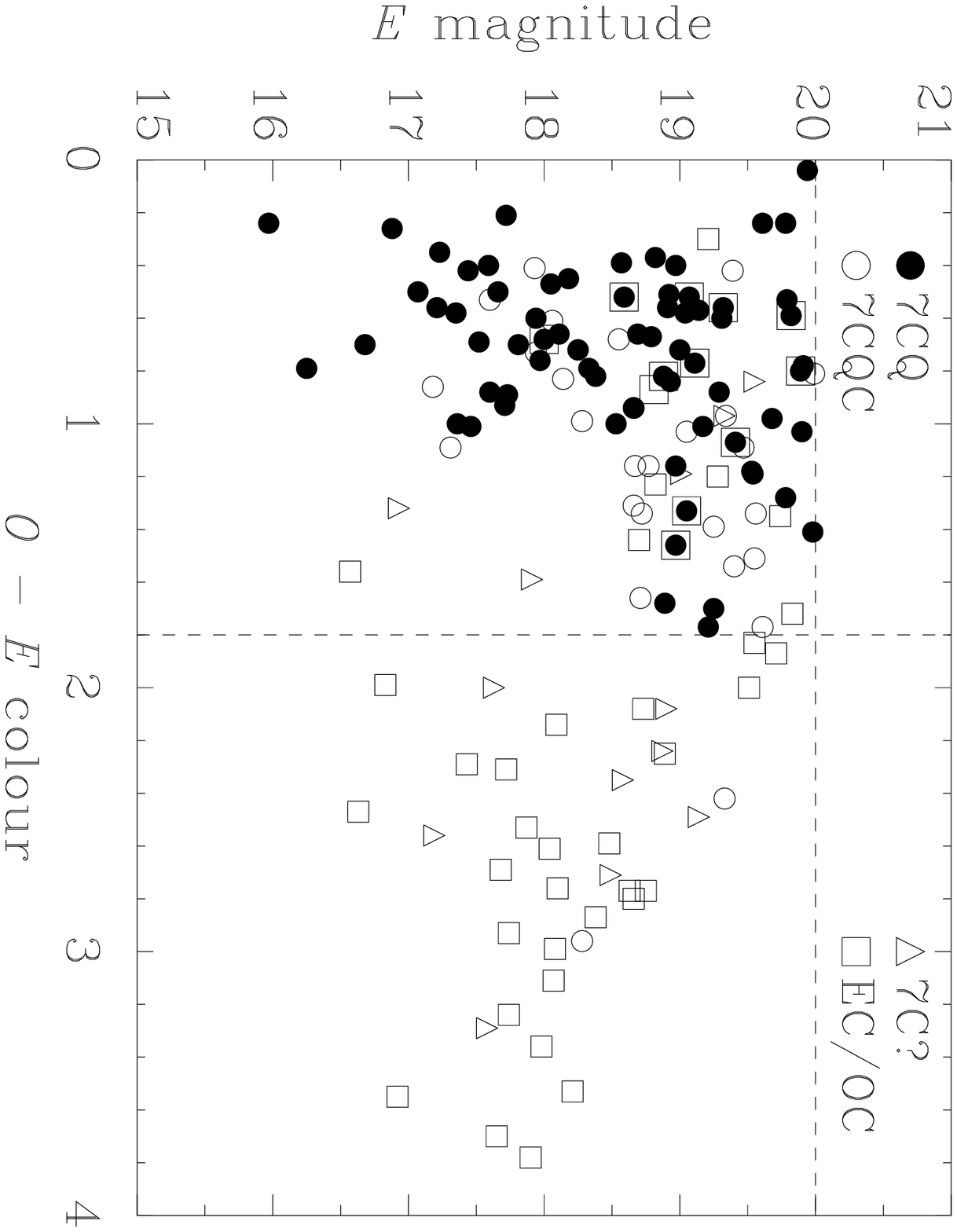}}
\end{picture}
\end{center}
{\caption[junk]{\label{fig:eo} 
The APM $E$ magnitude plotted against the $O-E$ colour
for the following categories of object: filled circles,
confirmed quasars (category `Q' in Tables 1 \& 2);
open circles and open triangles, candidate quasars which
are secure optical identifications (category `QC')
or insecure optical identifications (category `?')
respectively; open squares, objects classified as 
extended on one of the POSS-I plates (a union of the
EC1, EC2, OC0, OC1, OC2 categories defined in Table 1).
The dashed lines show the
effective $E \approx 20$ and $O-E \approx 1.8$ limits of the 7CQ sample, 
although we argue in Section 5.2 that only the
former leads to a significant loss of radio-loud quasars from the sample. 
These lines divide the plot into quadrants: objects detected only on the 
$E$ plate would fill out the top region of the bottom-right quadrant; 
and objects detected only on the
$O$ plate would lie in the top-left quadrant.
In the bottom-left quadrant there are
25 `QC' objects, 5 `?' objects and 7 `EC/OC' 
objects, and therefore 37 unconfirmed quasar
candidates remaining in the 7CQ sample. 
}}
\end{figure*}

The confirmed quasars all have colours bluer than
$(O - E) \approx 1.8$ as do 25 of the 27 objects in the `QC'
category. Only 5 of the 13 objects which lack VLA/optical follow-up 
(category `?') have colours bluer than this limit, so we can achieve a good level of formal completeness simply by 
adding this colour selection criterion. We will henceforth refer to the 25 blue `QC' objects
plus the other 11 blue objects (5 in category `?' and 6 
in one of the `EC/OC' categories) 
as the 36 unconfirmed quasar candidates.   
We shall argue in this section that this
additional colour criterion excludes few if any quasars. 

First, we will discuss the objects which the
APM analysis finds to be optically resolved on 
either or both of the plates. Ideally this
procedure should reject
quasar-like objects only if they are of low optical luminosity and lie at low redshift ($z \ll 0.5$) 
because only then can an optically-extended host galaxy contribute significantly to the
light. In this case the only quasars to be missed would be those whose
optical luminosities are lower than the luminosity of a giant galaxy: these 
systems are normally termed N-galaxies, or broad-line radio galaxies,
rather than quasars (Peacock, Miller \& Longair 1986; Willott et al. 1998b). However,
the existence of 11 spectroscopically-confirmed quasars in Table 1 (and in
Fig.\ 6) which the APM analysis categorised as not having simple unresolved optical structure on  
one of the plates indicates that some quasars would have been missed if the APM
classification had remained unquestioned. The origin of this problem is the breakdown of the APM classification
algorithm, particularly (as illustrated by the faint $E$ magnitudes of the
quasars `resolved' by the APM analysis) for objects near the magnitude limits of the POSS-I plates.
We thus attempted where possible to obtain VLA and spectroscopic follow-up of the fainter objects
in which APM classification was in doubt; only a few faint blue candidates were not followed up.
The majority of the objects in the `EC/OC' categories are resolved on the $E$ plate and are bright
enough in $E$ that the APM classification is likely to be reliable; they
have red $O-E$ colours just as one expects for low-redshift objects in 
which stellar light dominates. It is also possible that some genuinely point-like quasars are
spuriously classified as extended by the APM analysis; evidence that this 
can happen is provided by the source 0955+425 in Table 2 which was identified
as point-like by the first version of the APM classification analysis (see Section 2)
but which was classified as extended on both POSS-I plates by the second version. The number of cases
which have been missed completely is likely to be very small given that it requires
inaccurate classification on both plates, and by both versions of the APM analysis; these
hypothetical objects would be likely to have unusual properties (for example being projected 
close to a foreground star or galaxy) which would render them difficult cases for any
automated classification.

Second, we consider the 13 objects excluded from any follow-up (either with the VLA or spectroscopically),
namely those in category `?' in Tables 1 and 2. These were excluded from follow-up 
either because of large 7C angular size (1 case),
red optical colour (5 cases), or for random reasons (7 cases). We have already argued that 
excluding either red objects or large angular size radio sources is unlikely to reject quasars.
Since the remaining objects were excluded randomly, these exclusions should not bias the 
sample significantly.

Although we have followed up stellar objects regardless of colour from four of the 
five POSS-I plate pairs we have not yet found any 7C quasars with red colours
[$(O - E) > 1.8$]. Inspection of Fig.\ 6 shows that only two 7CQ objects are optically
unresolved, have red colours ($O-E > 1.8$) and currently lack spectroscopic follow-up.
Even if both of these are red quasars, they would constitute only a small ($\ll 10$ per cent) fraction of the population.
This is a direct consequence of two effects: first, 
the low fraction of lightly-reddened and/or intrinsically red quasars in samples selected at low 
radio frequency (Willott et al., 1998a, and in prep.); and second,
the tiny areal density expected for quasars at such high redshifts that their
Ly$\alpha$ line becomes redshifted beyond the peak in the $O-$plate sensitivity.

\subsection{Concluding remarks}

The discussion in Sections 5.1 and 5.2 highlights an extremely important problem
which is an ineluctable consequence of any survey for radio quasars which 
employs the POSS-I plates as the means of making optical identifications.
The depth of these plates is insufficient to identify all the quasars associated with a sample of 7C radio sources. 
Indeed, the majority of the radio-loud quasars above the $S_{151}$ limit will 
probably lie at magnitudes fainter than the POSS-I limits. The most direct indication of this comes from the 
optical magnitude distribution of the quasars, predominantly SSQs, in the 7C Redshift Survey (Willott et al. 1998b): 
in this survey optical spectra were taken of {\it all} objects with $S_{151} > 0.5 ~ \rm Jy$ in two 
patches of sky together covering
$0.013 ~ \rm sr$, and secure redshifts are now available for $\approx 90$ per cent of the sample 
(Blundell et al. 1998; Rawlings et al. 1998; Willott et al. 1998a). Just over 50 per cent 
of the quasars in this $S_{151} > 0.5 ~ \rm Jy$ sample are bright enough optically to 
be seen on either the $O$ or $E$ POSS-I plate. 
The correlation between the
radio and optical luminosities of SSQs (Serjeant et al. 1998) implies that 
an even lower fraction of quasars will appear on the POSS-I plates at fainter $S_{151}$ limits.
The 7CQ sample will therefore find only the
optically-brightest SSQs at any redshift, and 
is likely to miss $\approx 60$ per cent of the total number of SSQs
(Willott et al. 1998b). All analyses based 
on the 7CQ sample must take careful account of this effect. 
This means that, for example, a proper assessment of the quasar fraction
as a function of $S_{151}$ is beyond the scope of the present paper. 
A {\it lower limit} on the quasar fraction ($QF$) in the whole 7CQ sample is given by the
ratio of the total number of confirmed (category `Q') quasars (70) to the
total number of 7C radio sources (with $S_{151} > 0.1 ~ \rm Jy$) in the
region covered by 7C/APM cross-match (2409 sources),
namely $QF > 3$ per cent. Correcting for the unfinished spectroscopic follow-up and
various other sources of incompleteness (chiefly the expectation that more than half of the
radio-loud quasars fall below the magnitude limits of the POSS-I plates) 
could eventually yield a $QF \sim 10$ per cent. This may
prove to be significantly lower than the $QF \sim 20-30$ per cent seen 
in low-frequency radio samples selected at higher limiting values of $S_{151}$ 
(e.g. the 3C and 7C samples; Willott et al. 1998b).

Finally we note that the problem of insufficient plate depth may severely compromise other
SSQ surveys based on POSS-I IDs, e.g. the B3-VLA quasar sample (Vigotti et al. 1997).
Surveys which utilise deeper optical plates (e.g. UK Schmidt plates with a limiting magnitude of   
$B_{J} \approx 22.5$), and which have brighter $S_{151}$ selection limits --- and we are thinking here 
specifically  of the Molonglo/APM survey (MAQS; Serjeant et al. 1998, and in prep.) and the Molonglo Quasar 
Survey (MQS; Kapahi et al. 1998 and refs. therein) -- should be far less prone to this problem. The problems introduced for
SSQ surveys having dual radio and optical selection criteria are addressed quantitatively 
by Willott et al. (1998b).

\section*{Acknowledgements}

We are particularly grateful to Paul Alexander, Guy Pooley, 
Richard Saunders, Stephen Serjeant and Peter Warner 
who helped with various aspects of this work. We thank
the anonymous referee for some extremely useful suggestions.
We also thank the staff of the VLA. The
National Radio Astronomy Observatory is operated by Associated
Universities, Inc., under co-operative agreement with the National
Science Foundation. This
research has made use of the NASA/IPAC Extragalactic Database, which
is operated by the Jet Propulsion Laboratory, Caltech, under contract
with the National Aeronautics and Space Administration.

\end{document}